\documentclass[12pt, draftclsnofoot, onecolumn]{IEEEtran}

\usepackage{CJK}
\usepackage{color, soul, framed}
\usepackage{algorithm}
\usepackage{algorithmic}
\usepackage{amsmath}
\usepackage{amsthm}
\usepackage{amssymb}
\usepackage{psfrag}
\usepackage{graphicx}
\usepackage{epstopdf}
\usepackage{multirow}
\usepackage{stfloats}
\usepackage{cite}
\usepackage{subcaption}
\usepackage{color}
\usepackage[normalem]{ulem}
\usepackage{arydshln}
\usepackage{textcomp}
\usepackage{enumerate}
\usepackage{bbm}
\usepackage{xcolor}
\usepackage{xpatch}
\usepackage{array}
\usepackage{makecell}
\usepackage{multirow}
\usepackage{suffix}
\usepackage{mathtools}
\usepackage{diagbox}
\allowdisplaybreaks[4]
\DeclarePairedDelimiterX\MeijerM[3]{\lparen}{\rparen}%
{\begin{smallmatrix}#1 \\ #2\end{smallmatrix}\delimsize\vert\,#3}

\newcommand\MeijerG[8][]{%
  G^{\,#2,#3}_{#4,#5}\MeijerM[#1]{#6}{#7}{#8}}

\WithSuffix\newcommand\MeijerG*[7]{%
  G^{\,#1,#2}_{#3,#4}\MeijerM*{#5}{#6}{#7}}

\newtheorem{theorem}{\bf{Theorem}}
\newtheorem{lemma}{\bf{Lemma}}
\newtheorem{proposition}{\bf{Proposition}}

\newtheorem{remark}{\bf{Remark}}

\title{Multiple Access Design for Symbiotic Radios: Facilitating Massive IoT Connections \\ with Cellular Networks}
\author{Jun Wang, Xiangyu Ding, Qianqian Zhang, and Ying-Chang Liang, \emph{Fellow, IEEE}
\thanks{
Part of this work was presented in IEEE Globecom 2022\cite{wang2022multiple}. J. Wang, X. Ding and Q. Zhang are with the National Key Laboratory of Science and Technology on Communications, University of Electronic Science and Technology of China (UESTC), Chengdu 611731, China (e-mail: junwang@std.uestc.edu.cn, 202021220401@std.uestc.edu.cn, qqzhang\_kite@163.com). Y.-C. Liang is with the Center for Intelligent Networking and Communications (CINC), University of Electronic Science and Technology of China (UESTC), Chengdu 611731, China (e-mail: liangyc@ieee.org).
}
}

\begin{document}
 \maketitle

\vspace{-1.8cm}
\begin{abstract}
Symbiotic radio (SR) has emerged as a spectrum- and energy-efficient paradigm to support massive Internet of Things (IoT) connections. Two multiple access schemes are proposed in this paper to facilitate the massive IoT connections using the cellular network based on the SR technique, namely, the simultaneous access (SA) scheme and the selection diversity access (SDA) scheme. In the SA scheme, the base station (BS) transmits information to the receiver while multiple IoT devices transmit their information simultaneously by passively backscattering the BS signal to the receiver, while in the SDA scheme, only the IoT device with the strongest backscatter link transmits information to the receiver. In both of the schemes, the receiver jointly decodes the information from the BS and the IoT devices. To evaluate the above two schemes, in this paper, we have derived the closed-form expressions of the ergodic rates and the outage probabilities for the cellular and IoT transmissions. Finally, numerical results are provided to verify the theoretical analysis and compare the two proposed multiple access schemes. When the number of IoT devices is small, the SDA scheme is more appealing since it can significantly reduce the computational complexity while achieving equivalent performance to the SA scheme. When the number of IoT devices is large, the SA scheme is preferable since it guarantees a significantly better rate performance and a lower outage probability.


\end{abstract}

\section{Introduction}
The sixth generation (6G) wireless network is envisioned to realize ubiquitous communications to support a fully connected and intelligent digital world, which spurs an explosive growth in Internet of Things (IoT) connections~\cite{matti2019key,xiaohutowards,guo2021enabling,nguyen20216g}. According to the Ericsson mobility report, there will be around 30 billion IoT connections by 2027~\cite{Ericsson2021}. %The connectivity density will achieve $10^7$ devices per square kilometer~\cite{nguyen20216g}.
These massive IoT connections put significant strain on wireless networks due to the restricted spectrum and increased energy demand. Recently, a novel technique, called symbiotic radio (SR), has attracted significant attention to support IoT connections using cellular networks due to its spectrum and energy mutualistic sharing feature~\cite{long2019symbiotic,liang2020symbiotic,liang2022symbiotic}. In the SR system, the IoT device transmits information by passively backscattering the cellular signal, and thus the cellular and IoT transmissions share the same spectrum and energy resources.
%Recently, a novel technique, called symbiotic radio (SR), has been proposed as a promising technique for improving spectrum efficiency and energy efficiency~\cite{long2019symbiotic,liang2020symbiotic,liang2021symbiotic}. In the SR system, the backscatter device (BD) transmits information by passively backscattering the cellular signal, and thus the cellular and BD transmissions share the same spectrum and energy resources.
%requiring no additional spectrum or energy. The cellular transmission provides a strong signal carrier for the BD's transmission.
In return, due to the collaboration between the cellular and IoT transmissions, the backscattered signal serves as a multipath component instead of interference, leading to an enhancement to the cellular transmission.
%. Furthermore, when compared with conventional ambient backscatter communication (AmBC) systems~\cite{van2018ambient,liu2013ambient,yang2018cooperative}, the direct link interference is eliminated in SR due to the collaboration between the cellular and BD transmissions and joint decoding performed at the receiver.
Therefore, both cellular and IoT transmissions can benefit from their coexistence in SR. %Enhanced SR systems, such as multiple-input multiple-output (MIMO) SR~\cite{wu2021beamforming}, non-orthogonal multiple access (NOMA)-assisted SR~\cite{zhang2019backscatter,elsayed2021noma,raza2022noma}, reconfigurable intelligent surface-assisted SR~\cite{xu2021reconfigurable,wu2021reconfigurable}, and millimeter wave-based SR~\cite{yang2021joint,wang2021intelligent}, have been formed by combining several cellular transmission technologies with the SR technique.
Due to the outstanding spectrum and energy mutualistic sharing properties, SR has been regarded as an enabling technology for massive IoT connections in 6G networks ~\cite{imoize20216g,yazar20206g,chen2020vision}.

Currently, most researches on SR focus on the single-IoT device scenario and analyze the corresponding system performance, such as capacity~\cite{liu2018backscatter,guo2021capacity,zhou2019ergodic,darsena2017modeling} and outage probability~\cite{zhao2018ambient,zhang2019backscatter}. However, when it comes to massive IoT device scenario such as smart factories and warehousing, existing performance analyses for the single-IoT device SR scenario cannot be simply extended since there will involve interference among IoT devices.
%Multiple IoT devices access scenarios, in which massive IoT devices need to transmit information, are common in reality, such as smart factories and warehousing.
%Because there is interference among IoT devices and the transceiver design is substantially more sophisticated, SR systems with multiple IoT devices differ significantly from single IoT device systems. Existing performance analyses for single IoT device systems~\cite{liu2018backscatter,zhao2018ambient,zhou2019ergodic} cannot be simply extended to the multi-IoT device scenario.
Various multiple access schemes have been investigated in the literature to enable massive IoT connections with the SR technique~\cite{yang2021energy,han2021design,liu2020symbiotic,chen2020stochastic,ding2021advantages,zhang2021minimum}. In particular, the time division multiple access (TDMA) scheme was proposed in~\cite{yang2021energy}, in which each IoT device is assigned a specific time slot and takes turns to transmit information in the assigned time slot. The transmit power, IoT devices' reflection coefficients, and time slot were optimized to maximize the system energy efficiency. When the IoT device with the potential highest throughput was allocated with the maximum allowed time for transmission, the maximum energy efficiency could be achieved. In~\cite{han2021design}, the code division multiple access (CDMA) scheme was proposed, whereas each IoT device independently precoded its information with a random spreading code.
%The asymptotic signal-to-interference-plus-noise ratio (SINR) was found to be independent of the specific random codes and only determined by the system load and statistical channel state information (CSI).
%The minimum signal-to-interference-plus-noise ratio (SINR) of the BDs was maximized by jointly designing the transmit power and BDs' reflection coefficients.
The proposed random coded-based scheme avoided the need for coordination among the IoT devices, which was appropriate for IoT devices with limited signal processing capability. Another CDMA-based multiple access scheme was proposed in~\cite{liu2020symbiotic}, which could avoid the interference among multiple IoT devices by applying the $\mu$code encoding mechanism, and could achieve an effective synchronous or asynchronous transmission. However, the proposed scheme was invalid when the access time between two IoT devices was greater than the code length. The spatial division multiple access (SDMA) scheme was proposed in~\cite{chen2020stochastic} and multiple antennas were required at the receiver to alleviate the direct-link interference and the interference among IoT devices through stochastic transceiver design. In~\cite{ding2021advantages}, the non-orthogonal multiple access (NOMA)-based multiple access scheme was investigated. The sum rate of the IoT devices was maximized by optimizing the IoT devices' reflection coefficients. It was demonstrated that even with a random resource allocation scheme, the NOMA-based scheme outperformed the orthogonal multiple access (OMA) scheme in terms of transmission rate. A hybrid access scheme incorporating NOMA and TDMA was studied in~\cite{zhang2021minimum}. Multiple IoT devices were classified into several clusters, and each cluster transmitted information during the corresponding time slot. Within the cluster, the IoT devices adopted the NOMA protocol to transmit information. By designing the transmit power, IoT devices' reflection coefficients and the time slot, the minimum throughput among all IoT devices was maximized. Despite the fact that several specific multiple access schemes for SR have been proposed, two issues remain to be investigated: 1) what is the relationship between the achievable rate and the number of the accessed IoT devices in SR systems; 2) whether it is better to serve multiple IoT devices simultaneously or only one specific device at a time.

To address the above two issues, in this paper, we propose two multiple access schemes for SR, namely, the simultaneous access (SA) scheme and the selection diversity access (SDA) scheme. In the SA scheme, the base station (BS) transmits information to the receiver, and multiple IoT devices backscatter their information simultaneously over the BS signal. The receiver decodes the information from the BS and IoT devices using joint decoding and successive interference cancellation (SIC) techniques. Since there is always an IoT device with a substantially stronger channel strength than the average, rather than allowing all the IoT devices to connect to the SR network, only the IoT device with the strongest backscatter link is chosen to transmit its information in the SDA scheme. The receiver decodes both the BS and the chosen IoT device's information. In the long run, each IoT device has the opportunity to be chosen and transmits its information to the receiver due to the random and independent fading nature. By exploiting multiple IoT devices and spatial diversity, the multiuser diversity gain can be realized. To evaluate the proposed multiple access schemes, we then derive closed-form expressions of the ergodic rates and outage probabilities for cellular and IoT transmissions. Based on the analyses and simulations, we discover that: 1) the SA scheme is more appealing when the number of IoT devices $K$ is large since it guarantees a significantly better rate performance and a lower outage probability; 2) the SDA scheme is superior when $K$ is small since the SDA scheme can reach 85 percent of the performance of the SA scheme while significantly reducing complexity; 3) the ergodic sum rate of the IoT transmission in the SA scheme scales with $\mathrm{ln} K$, while in the SDA scheme it scales with $\mathrm{ln}\left(\mathrm{ln} K\right)$; 4) in SDA scheme, the tail of the distribution with double fading is heavier than that with single fading, leading to a higher multiuser diversity gain and transmission rate, which implies that the backscatter link’s double fading effect can also be beneficial for transmission.

In a nutshell, the main contributions of this paper are summarized as follows:
\begin{itemize}
\item We propose two multiple access schemes for SR, namely, SA and SDA schemes, to facilitate massive IoT connections using the cellular networks. In the SA scheme, the BS transmits information to the receiver while multiple IoT devices transmit their information simultaneously by passively backscattering the BS signal. In the SDA scheme, only the IoT device with the strongest backscatter link can transmit information to the receiver.
\item For both multiple access schemes, we derive the closed-form expressions of ergodic rates for cellular and IoT transmissions, which reveal the relationship between the achievable rate and the number of the accessed IoT devices in SR systems. In addition, the closed-form expressions of outage probabilities are also provided.
\item We compare the performance of the proposed schemes and discuss their respective applicable scenarios. When the number of IoT devices is small, the SDA scheme is more appealing, and vice versa, the SA scheme is preferable.
\item We show that both cellular and IoT transmissions in the SDA scheme benefit from the double fading effect since a higher multiuser diversity gain can be guaranteed.
\end{itemize}

The rest of the paper is organized as follows. In Section~\ref{system model}, we introduce the symbiotic radio multiple access system, including system description and signal models for the SA and SDA schemes. In Section~\ref{sec_rate}, the theoretical expressions in terms of the ergodic rate for the proposed schemes are derived. In Section~\ref{sec_outage}, we derive the outage probabilities of the proposed schemes. The numerical results are presented in Section~\ref{sec-simulation} to verify the accuracy of the theoretical results and compare the performance of the proposed multiple access schemes. Finally, the paper is concluded in Section~\ref{sec-conclusion}.

The notations involved in the paper are as follows. $|\cdot|$ denotes the absolute value. ${\cal{N}}(\mu,\sigma^2)$ and ${\cal{CN}}(\mu,\sigma^2)$ denote the real Gaussian distribution and the complex Gaussian distribution with mean $\mu$ and variance $\sigma^2$. $f_X(\cdot)$, $F_X(\cdot)$, $\mathbb{E}(X)$ and $\mathbb{V}(X)$ denote the probability density function (PDF), cumulative distribution function (CDF), expectation and variance of random variable (RV) $X$, respectively. $\mathrm{Re}\{x\}$ and $\mathrm{Im}\{x\}$ represent the real part and the imaginary part of variable $x$, respectively.
%Although several works have studied the multi-BD SR system, the theoretical analysis of the system performance is absent. We note that the multiple access scheme for multi-BD AmBC system has been studied and provides us basics for the performance analysis of multi-BD SR system.

%To the authors' best knowledge, the performance of the multi-BD SR multiple access system has not been investigated yet.
%potential benefits offered by SR technique, significant research efforts have been devoted to investigate the performance of SR systems.

%\subsubsection{Related Works}
%The multiple access for multi-BD AmBC system has been investigated in the literature.
%\subsubsection{Organization}

\section{System Model}\label{system model}
In this section, we will present the system model for symbiotic radio multiple access systems, including the system description and signal models for SA and SDA schemes.
\subsection{System Description}
\begin{figure}[!t]
\centering
\includegraphics[width=0.7\columnwidth] {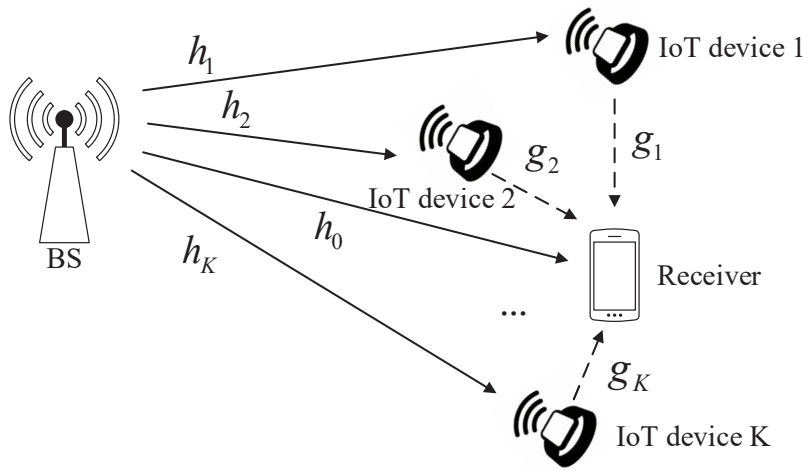}\vspace{-0.3cm}
\caption{System model for symbiotic radio multiple access systems.}
\label{fig:Fig1}
\vspace{-0.4cm}
\end{figure}

\begin{figure}[!t]
\centering
\includegraphics[width=0.9\columnwidth] {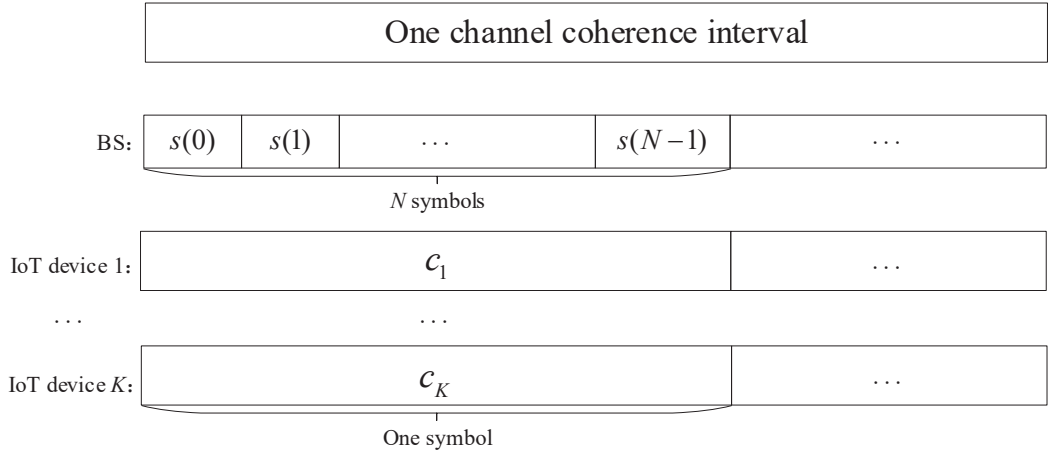}\vspace{-0.3cm}
\caption{An illustration of the timeline for transmitted symbols from the BS and IoT devices.}
\label{fig:Fig_transmission_frame}
\vspace{-0.4cm}
\end{figure}

Fig. \ref{fig:Fig1} illustrates the symbiotic radio multiple access system, consisting of a BS, a receiver, and multiple active IoT devices around the receiver$\footnote{Two typical application scenarios for the proposed system are healthcare, and logistics and warehousing, in which multiple IoT devices are deployed close to the receiver and the time delay between the direct link and backscatter link is thus negligible.}$. The BS transmits information to the receiver, while multiple IoT devices transmit their information over the signal received from the BS. The receiver needs to decode the information from the BS and the IoT devices. The BS, IoT device, and receiver are assumed to be equipped with a single antenna. We consider a block fading channel model in this paper. As shown in Fig. \ref{fig:Fig1}, $h_0$, $h_k$, and $g_k$ denote the channel responses from the BS to the receiver, from the BS to $k$-th IoT device, and from the $k$-th IoT device to the receiver, respectively. We assume that the symbol period of the IoT device is $N$ times that of the BS, as illustrated in Fig.~\ref{fig:Fig_transmission_frame}. The transmitted symbols from the BS and the $k$-th IoT device are denoted by $s(n)\sim {\cal{CN}}\left(0,1\right)$ and $c_k\sim {\cal{CN}}\left(0,1\right)$, respectively.

\begin{figure}[!t]
\centering
\includegraphics[width=0.95\columnwidth] {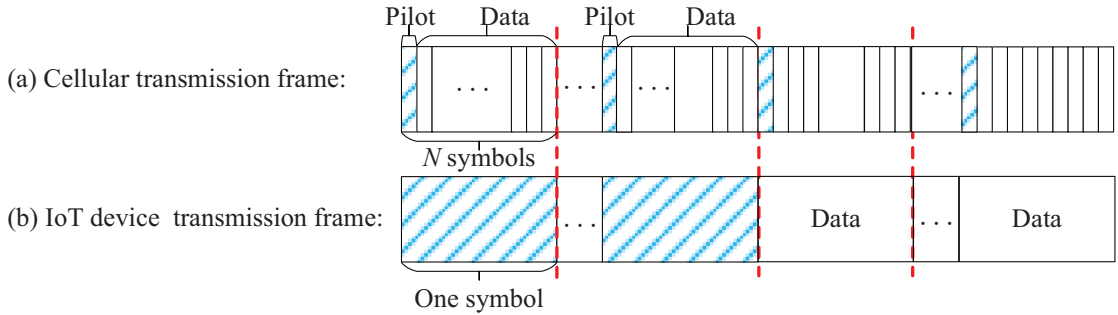}\vspace{-0.3cm}
\caption{Frame structure in symbiotic radio multiple access systems.}
\label{fig:Fig_frame}
\vspace{-0.4cm}
\end{figure}

The frame structure is presented in Fig.~\ref{fig:Fig_frame}. The BS and IoT devices first send pilots, and the receiver can perform channel estimation to obtain the channel state information (CSI) and feedback to the BS. The data transmission is then carried out. In the following, the performance analyses are given under the perfect CSI assumption$\footnote{In this paper, we have assumed perfect CSI to investigate the best achievable performance. The performance under imperfect CSI is worth further investigation by refering to existing analytical works~\cite{mishra2019sum,chen2015interference}.}$. In the SA scheme, multiple IoT devices backscatter their information to the receiver, while in the SDA scheme, only the IoT device with the strongest backscatter link can transmit information. In the following, we introduce the signal models for these two schemes.

\subsection{Signal Model for Simultaneous Access (SA) Scheme}
In the SA scheme, the receiver receives two types of signals: the direct link signal from the BS and the backscatter link signal from $K$ IoT devices. The signal received by the receiver can be written as
\begin{align}
    y(n)=\sqrt{p}h_0s(n)+\sqrt{p}\sum_{k=1}^K\alpha_kg_kh_kc_ks(n)+u(n),
\end{align}
where $p$ is the BS transmit power and $\alpha_k$ is the reflection coefficient of the $k$-th IoT device. %Without loss of generality, we assume that $\alpha_k=\alpha, \forall k$.
$u(n)$ is the additive complex Gaussian noise that follows the distribution of ${\cal{CN}}(0,\sigma^2)$. The receiver first decodes the information from the BS, and then decodes the information from the IoT devices by using successive interference cancellation (SIC) technique. The signal-to-noise-ratio (SNR) for $s(n)$ is given by~\cite{liu2018backscatter}
\begin{align}
    \gamma_s^{SA}=\frac{p|h_0+ \sum_{k=1}^K\alpha_k g_kh_kc_k|^2}{\sigma^2}.\label{gammas}
\end{align}
The expression of $\gamma_s^{SA}$ involves $\left\{c_k\right\}_{k=1}^K$, which changes faster as compared to the channel variation. The achievable rate of cellular transmission is thus given by~\cite{shin2003capacity,jeon2011bounds}
\begin{align}
    R_s^{SA}=\mathbb{E}_{\left\{c_k\right\}_{k=1}^K}\left[\mathrm{log}_2\left(1+\gamma_s^{SA}\right)\right],\label{original}
\end{align}
where the expectation is taken over the random variables $\left\{c_k\right\}_{k=1}^K$. It is rather challenging to obtain the closed-form expression for $R_s^{SA}$. To obtain analytical insights, we consider the asymptotic case with high SNR in the following proposition.
\begin{proposition}\label{theorem1}
The achievable rate in high SNR regimes is given by
\begin{align}
    R_s^{SA} \approx \mathrm{log}_2 \left(\frac{p|h_0|^2}{\sigma^2}\right)-\mathrm{Ei}\left(-\frac{|h_0|^2}{\sum_{k=1}^K|\alpha_k g_kh_k|^2}\right)\mathrm{log}_2 e,\label{Rs_original_highSNR}
\end{align}
where $\mathrm{Ei}(x)=\int_{-\infty}^{x}\frac{e^u}{u}du$ is the exponential integral.
\end{proposition}
\begin{IEEEproof}
The details are given in Appendix \ref{proof of gamma_s for high SNR}.
\end{IEEEproof}

After decoding $s(n)$, the receiver removes the direct link interference and then decodes the information of IoT devices by using SIC technique. Note that the symbol period of the IoT device is $N$ times that of the BS, the signal-to-interference-plus-noise ratio (SINR) for $c_j$ via maximal ratio combining (MRC) can be given by
\begin{align}
    \gamma_{c_j}^{SA}=\frac{Np|\alpha_jg_jh_j|^2}{Np\sum_{k=j+1}^K|\alpha_k g_kh_k|^2+\sigma^2}.
\end{align}
The sum rate of $K$ IoT devices is thus given by
\begin{align}
    R_{c}^{SA}=\frac{1}{N}\mathrm{log}_2\left(1+\frac{Np\sum_{k=1}^K|\alpha_k g_kh_k|^2}{\sigma^2}\right).\label{c_rate}
\end{align}
\begin{remark}
Since the function $-\mathrm{Ei}(-x)$ decreases as $x$ increases, according to (\ref{Rs_original_highSNR}), we obtain that $R_s$ monotonically increases as the number of IoT devices $K$ increases. The more IoT devices are connected to the SR network, the higher cellular transmission rate can be achieved. Meanwhile, the sum rate of IoT devices also increases as $K$ increases, as indicated by (\ref{c_rate}).
\end{remark}
\subsection{Signal Model for Selection Diversity Access (SDA) Scheme}
%\begin{figure}[!t]
%\centering
%\includegraphics[width=0.95\columnwidth] {Visio_frame_structure_SDA.eps}\vspace{-0.3cm}
%\caption{\textcolor{blue}{Frame structure for SDA scheme in symbiotic radio multiple access systems.}}
%\label{fig:Fig_frame_SDA}
%\vspace{-0.4cm}
%\end{figure}
There is always an IoT device with a substantially stronger channel strength than the average~\cite{viswanath2002opportunistic}. In the SDA scheme, instead of allowing all of the IoT devices to access the network simultaneously, the BS chooses the IoT device with the strongest backscatter link to transmit its information based on the feedback CSI, while other IoT devices are informed to stay idle. In this way, the computational complexity at the receiver can be reduced and interference among the IoT devices can be avoided$\footnote{
For the more general case in which not all the IoT devices are active (the IoT device with the strongest backscatter link may not be available), we can first perform activity detection and then apply the SDA scheme for the active IoT devices, or utilize the general order selection scheme ~\cite{al2020performance,mishra2019sum} with order statistics theory~\cite{david2004order}.}$. In this configuration, all the IoT devices receive the signal from the BS simultaneously, but only the chosen IoT device denoted as the $l$-th IoT device, can transmit $c_l$ to the receiver. The received signal at the receiver is given by
\begin{align}
    y(n)=\sqrt{p}h_0s(n)+\sqrt{p}\alpha_l g_lh_lc_ls(n)+u(n),
\end{align}
where $l=\mathrm{arg}\underset{k}{\max}|\alpha_kg_kh_k|^2$. The receiver first decodes $s(n)$ and then decodes $c_l$. The SNR for $s(n)$ is given by
\begin{align}
    \gamma_s^{SDA}=\frac{p|h_0+\alpha_l g_lh_lc_l|^2}{\sigma^2}.
\end{align}
Similar to Proposition~\ref{theorem1}, we obtain the achievable rate of cellular transmissions as
\begin{align}
    R_s^{SDA}&=\mathrm{log}_2 \left(\frac{p|h_0|^2}{\sigma^2}\right)-\mathrm{Ei}\left(-\frac{|h_0|^2}{ \underset{k}{\max}|\alpha_k g_kh_k|^2}\right)\mathrm{log}_2 e.
\end{align}
For given $s(n)$, the SNR for $c_l$ via MRC is expressed as
\begin{align}
    \gamma_{c_l}^{SDA}=\frac{Np\underset{k}{\max}|\alpha_k g_kh_k|^2}{\sigma^2}.
\end{align}
The corresponding IoT transmission rate is given by
\begin{align}
    R_c^{SDA}&=\frac{1}{N}\mathrm{log}_2\left(1+\frac{Np\underset{k}{\max}|\alpha_k g_kh_k|^2}{\sigma^2}\right).
\end{align}
%\begin{remark}

%\end{remark}
\section{Ergodic Rate}\label{sec_rate}
In this section, we analyze the ergodic rates to eliminate the influence of the channel fading and compare the ergodic rate of cellular transmission as well as the ergodic sum rate of IoT transmission in proposed two multiple access schemes to obtain scheme selection insights. We assume that the channels $h_0$, $h_k$ and $g_k$ follow Gaussian distribution, i.e., $h_0\sim{\cal{CN}}\left(0,\lambda_0\right)$, $h_k \sim {\cal{CN}}\left(0,\lambda_h\right)$, and $g_k\sim {\cal{CN}}\left(0,\lambda_g\right)$~\cite{zhou2019ergodic,jin2014ergodic}. Denote $\boldsymbol{h}=\left\{h_0,h_1,\cdots,h_K,g_1,\cdots,g_K\right\}$, $Z_k=|g_kh_k|^2$, $\Lambda=\sum_{k=1}^K|g_kh_k|^2$, $Z=\underset{k}{\max}\ |g_kh_k|^2$, and $\lambda=\lambda_g \lambda_h$. To simplify the derivation, herein after we assume $\alpha_k=\alpha,\forall k$.%Since the exact distribution of $\Lambda$ is complicated, we utilize Gaussian distribution to approximate the distribution of $\Lambda$. The central limit theorem suggests that $\Lambda$ follows the Gaussian distribution ${\cal{N}}\left(K\lambda,3K\lambda^2\right)$, whose detailed proof is presented in Appendix \ref{proof of gaussian}. %For small $K$, the Gaussian approximation may not work well and we use Gamma distribution $Gamma\left(\beta_1,\beta_2\right)$ to approximate the distribution of $\Lambda$, similarly to \cite{zhao2020backscatter}. The parameters $\beta_1$ and $\beta_2$ can be obtained with the expectation and variance of $\Lambda$ and we get
%\begin{align}
  %  \beta_1\beta_2&=K\lambda,\\
  %  \beta_1\beta_2^2&=3K\lambda^2.
%\end{align}
%Thus, we have
%\begin{align}
  %  \beta_1=\frac{K}{3},\beta_2=3\lambda.
%\end{align}

%from the central limit theorem, we can obtain that

\subsection{Ergodic Rate Analysis for Simultaneous Access Scheme}
\subsubsection{Ergodic Rate of Cellular Transmission}
The ergodic rate for $s(n)$ can be expressed as
\begin{align}
    \mathbb{E}_{\boldsymbol{h}}\left[R_s^{SA}\right]=\mathbb{E}_{\boldsymbol{h_0}}\left[\mathrm{log}_2 \left(\frac{p|h_0|^2}{\sigma^2}\right)\right]-\mathbb{E}_{\boldsymbol{h_0},\Lambda}\left[\mathrm{Ei}\left(-\frac{|h_0|^2}{\alpha^2 \Lambda}\right)\right]\mathrm{log}_2 e.\label{eq_Eh}
\end{align}
The closed-form expression of the ergodic rate of cellular transmission can be obtained in the following theorem.
\begin{theorem}
The ergodic rate of cellular transmission is given by
\begin{align}
    \mathbb{E}_{\boldsymbol{h}}\left[R_s^{SA}\right]=&\mathrm{log}_2\left(\frac{p}{\sigma^2}\right)+\bigg[\mathrm{ln}\left(1+\frac{K\alpha^2\lambda}{\lambda_0}\right)-\frac{3K\lambda^2\alpha^4}{2\left(\lambda_0+K\alpha^2\lambda\right)^2}-E_0\bigg]\mathrm{log}_2 e,\label{s_ergodic}
\end{align}
where $E_0$ is the Euler-Mascheroni constant.
\begin{IEEEproof}
The details are given in Appendix~\ref{ergodic_rate_s_SA}.
\end{IEEEproof}
\end{theorem}
The high SNR slope shows the influence of channel parameters on the ergodic rate. The slope in high SNR regimes is defined as: $S=\underset{p\rightarrow \infty}{\mathrm{lim}}\frac{R\left(p\right)}{10\mathrm{log}_{10}p}$~\cite{zhang2019backscatter}.
The slope in terms of ergodic rate of cellular transmission is thus given by $S_{s,SA}=\frac{\mathrm{log}_2(10)}{10}\approx 0.332$.

\subsubsection{Ergodic Sum Rate of IoT Transmission}
The ergodic sum rate of IoT transmission can be given by
\begin{align}
    &\mathbb{E}_{\boldsymbol{h}}\left[R_{c}^{SA}\right]
    =\mathbb{E}_{\Lambda}\left[\frac{1}{N}\mathrm{log}_2\left(1+\frac{Np\alpha^2\Lambda}{\sigma^2}\right)\right].
\end{align}
 Using the second-order Taylor expansion of $\mathrm{ln}(1+x)$ with a Gaussian
distribution input, i.e., $\mathbb{E}\left[\mathrm{ln}\left(1+x\right)\right]\approx \mathrm{ln}\left(1+\mathbb{E}\left[x\right]\right)-\frac{\mathbb{V}[x]}{2\left(1+\mathbb{E}\left[x\right]\right)^2}$, we obtain the ergodic rate approximation as
\begin{align}
    &\mathbb{E}_{\boldsymbol{h}}\left[R_{c}^{SA}\right]\approx \frac{1}{N}\left[\mathrm{ln}\left(1+\frac{K\lambda Np\alpha^2}{\sigma^2}\right)-\frac{3K\lambda^2N^2p^2\alpha^4}{2\left(\sigma^2+K\lambda Np\alpha^2\right)^2}\right]\mathrm{log}_2 e.\label{c_ergodic}
\end{align}
The slope in terms of ergodic sum rate of IoT transmission is given by $S_{c,SA}=\frac{\mathrm{log}_2(10)}{10N}\approx \frac{0.332}{N}$.
\begin{remark}
Based on previous expressions, we observe that both $R_s^{SA}$ and $R_c^{SA}$ both scales with $\mathrm{ln}(K)$. Meanwhile, $R_c^{SA}$ scales with $\frac{1}{N}\mathrm{ln}(N)$, which is consistent with the results of the single IoT device scenario in~\cite{zhou2019ergodic}.
\end{remark}

%When $K \text{mod}\,3=0$, $\beta_1-1$ is a non-negative integer. Denote $\xi_1=\frac{\lambda_0}{\alpha^2\beta_2}$. From \cite{gradshteyn2014table}, we have
%\begin{align}
    %&\int_0^{\infty}\text{ln}\left(1+\frac{\alpha^2\beta_2 t}{\lambda_0}\right)t^{\beta_1-1}e^{-t}dt\nonumber\\
   % &=\sum_{\mu=0}^{\beta_1-1}\frac{\left(\beta_1-1\right) !}{\left(\beta_1-1-\mu\right)!}\Bigg[-\left(-\xi_1\right)^{\beta_1-1-\mu}e^{\xi_1}\text{Ei}\left(-\xi_1\right)\nonumber\\
    %&+\sum_{k=1}^{\beta_1-1-\mu}\left(k-1\right)!\left(-\xi_1\right)^{\beta_1-1-\mu-k}\Bigg]\triangleq v\left(\beta_1,\xi_1\right).\label{integration_ln}
%\end{align}
%By substituting (\ref{expectation_first}), (\ref{E_gamma}) and (\ref{integration_ln}) into (\ref{eq_Eh}), we obtain the ergodic rate from BS to receiver under Gamma approximation as
%\begin{align}
    % \mathbb{E}_{\boldsymbol{h}}\left[R_s\right]=\text{log}_2\left(\frac{p}{\sigma^2}\right)+\text{log}_2 e\left[\frac{v\left(\beta_1,\xi_1\right)}{\Gamma\left(\beta_1\right)}-E_0\right].
%\end{align}
%When $K \text{mod}\,3\neq 0$,

\subsection{Ergodic Rate Analysis for Selection Diversity Access Scheme}\label{ergodic_SDA}
In this subsection, the ergodic rates of the multiple IoT devices selection diversity access scheme are investigated. The closed-form expressions of ergodic rates of cellular and IoT transmissions are derived by using probabilistic analysis and approximation methods. Moreover, the asymptotic ergodic sum rate analysis for IoT transmission is also provided to provide more insights by using extreme value theory. The following lemma is used to obtain the closed-form expressions of ergodic rates.
\begin{lemma}
  For the function $f(x)=\mathrm{ln}\left(1+\epsilon x\right)\left(1-2\zeta\sqrt{ x}K_1\left(2\zeta\sqrt{ x}\right)\right)^{K-1}K_0\left(2\zeta\sqrt{ x}\right)$, $\underset{x \rightarrow \infty}{\mathrm{lim}}f(x)=0$, where $\epsilon>0$, $\zeta>0$, $K_0(x), K_1(x)$ are the modified Bessel functions of the second kind.\label{pro3}
\end{lemma}
\begin{IEEEproof}
The details are given in Appendix~\ref{proof of lemma1}.
\end{IEEEproof}
%, including exact and asymptotic ergodic rate. The exact ergodic rate analysis provides the closed-form expression by using probabilistic analysis and approximation methods while the asymptotic ergodic rate analysis provides more insights
%\subsubsection{Ergodic Rate of cellular transmission}
\subsubsection{Ergodic Rate of Cellular Transmission}
The ergodic rate for $s(n)$ can be given by
\begin{align}
    \mathbb{E}_{\boldsymbol{h}}\left[R_s^{SDA}\right]=\mathbb{E}_{\boldsymbol{h_0}}\left[\mathrm{log}_2 \left(\frac{p|h_0|^2}{\sigma^2}\right)\right]-\mathbb{E}_{\boldsymbol{h_0},Z}\left[\mathrm{Ei}\left(-\frac{|h_0|^2}{\alpha^2 Z}\right)\right]\mathrm{log}_2 e.
\end{align}
The closed-form expression of ergodic rate of cellular transmission can be obtained from the following theorem.
\begin{theorem}
The ergodic rate of cellular transmission is given by
\begin{align}
    \mathbb{E}_{\boldsymbol{h}}\left[R_s^{SDA}\right] \approx \mathrm{log}_2\left(\frac{p}{\sigma^2}\right)- E_0\mathrm{log}_2 e+\frac{KM_1\pi\mathrm{log}_2 e}{\lambda n}\sum_{i=1}^{n}l\left(\phi_i\right), \label{E_Rs_SDA}
\end{align}
where $M_1$ is a large constant, $\phi_i=\mathrm{cos}\left(\frac{2i-1}{2n}\pi\right)$, $n$ is a complexity-accuracy tradeoff parameter and $l(x)$ is written as
\begin{align}
l(x) &=\mathrm{ln}\left(1+\frac{\alpha^2 \left(M_1x+M_1\right)}{2\lambda_0}\right)\Bigg(1-2\sqrt{\frac{M_1x+M_1}{2\lambda}} K_1\left(2\sqrt{\frac{M_1x+M_1}{2\lambda}}\right)\Bigg)^{K-1}\nonumber\\
&\times K_0\left(2\sqrt{\frac{M_1x+M_1}{2\lambda}}\right)  \sqrt{1-x^2}.
\end{align}
\begin{IEEEproof}
The details are given in Appendix~\ref{proof of rate s for SDA}.
\end{IEEEproof}
\end{theorem}
For SDA scheme, the slope in terms of ergodic rate of cellular transmission can be given by $S_{s,SDA}\approx 0.332$.
\subsubsection{Exact Ergodic Rate of IoT Transmission}
The ergodic rate for $c_l$ can be given by
\begin{align}
    \mathbb{E}_{\boldsymbol{h}}\left[R_{c}^{SDA}\right]
    &=\mathbb{E}_Z\left[\frac{1}{N}\mathrm{log}_2\left(1+\frac{Np\alpha^2Z}{\sigma^2}\right)\right]=\int_0^{\infty}\frac{1}{N}\mathrm{log}_2\left(1+\frac{Np\alpha^2z}{\sigma^2}\right)f_Z(z)dz.
\end{align}
According to Lemma~\ref{pro3}, we transform the integral into
\begin{align}
    &\mathbb{E}_{\boldsymbol{h}}\left[R_{c}^{SDA}\right]\approx\int_0^{M_2}\frac{1}{N}\mathrm{log}_2\left(1+\frac{Np\alpha^2z}{\sigma^2}\right)\frac{2K}{\lambda} \left(1-2\sqrt{\frac{z}{\lambda}}K_1\left(2\sqrt{\frac{z}{\lambda}}\right)\right)^{K-1}K_0\left(2\sqrt{\frac{z}{\lambda}}\right)dz.
\end{align}
where $M_2$ is a large constant. Applying Gaussian-Chebyshev quadrature, we have
\begin{align}
    \mathbb{E}_{\boldsymbol{h}}\left[R_{c}^{SDA}\right]\approx \frac{KM_2\pi}{Nn\lambda}\sum_{i=1}^n m\left(\phi_i\right),\label{E_Rc_SDA}
\end{align}
where
\begin{align}
    m(x)&=\mathrm{log}_2\left(1+\frac{Np\alpha^2(M_2x+M_2)}{2\sigma^2}\right) \left(1-2\sqrt{\frac{M_2x+M_2}{2\lambda}}K_1\left(2\sqrt{\frac{M_2x+M_2}{2\lambda}}\right)\right)^{K-1}\nonumber\\
    &\times K_0\left(2\sqrt{\frac{M_2x+M_2}{2\lambda}}\right)\sqrt{1-x^2}.
\end{align}
\subsubsection{Asymptotic Ergodic Rate of IoT Transmission}
The exact ergodic rate formulation for IoT transmission, as seen in (\ref{E_Rc_SDA}), is complicated and lacks insights. As a result, in this part, we present an asymptotic ergodic sum rate analysis in order to gain a better understanding of the relationship between the ergodic rate and the number of IoT devices. The following lemma comes from the extreme value theory~\cite{viswanath2002opportunistic,song2006asymptotic,li2019capacity}.
\begin{lemma}
Let $\xi_1,\xi_2,\cdots,\xi_K$ be independent and identically distributed (i.i.d.) RVs with a common CDF $F(x)$ and PDF $f(x)$, which satisfies $F(x)\leq 1$ and $F''(x)$ exist for all $x$. Denote $\xi=\underset{i}{\mathrm{max}} \ \xi_i$. If
\begin{align}
    \underset{x\rightarrow \infty}{\mathrm{lim}} \frac{d}{dx}\left[\frac{1-F(x)}{f(x)}\right]=0,
\end{align}
then $\frac{\xi-a_K}{b_K}$ uniformly converges in distribution to a normalized Gumbel RV as $K \rightarrow \infty$, whose CDF, expectation, and variance are given by $e^{-e^{x}}$, $E_0$ and $\frac{\pi^2}{6}$, respectively. The normalizing constants $a_K$ and $b_K$ are respectively derived as
\begin{align}
   a_K&=F^{-1}\left(1-\frac{1}{K}\right),\\
   b_K&=F^{-1}\left(1-\frac{1}{Ke}\right)-F^{-1}\left(1-\frac{1}{K}\right),
\end{align}
where $F^{-1}(x)=\mathrm{inf}\left\{y:F(y)\geq x\right\}$.
\end{lemma}
By using the PDF and CDF of $Z_i$, we have
\begin{align}
    \underset{x\rightarrow \infty}{\mathrm{lim}} \frac{d}{dx}\left[\frac{1-F(x)}{f(x)}\right]
    &=\underset{x\rightarrow \infty}{\mathrm{lim}}-1-\frac{\left[1-F(x)\right]f'(x)}{f^2(x)}=-1+\underset{x\rightarrow \infty}{\mathrm{lim}}\left(\frac{K_1\left(2\sqrt{\frac{x}{\lambda}}\right)}{K_0\left(2\sqrt{\frac{x}{\lambda}}\right)}\right)^2.
\end{align}
Since $K_1(x)$ and $K_0(x)$ approach $\sqrt{\frac{\pi}{2x}}e^{-x}$ as $x\rightarrow \infty$, we obtain $\underset{x\rightarrow \infty}{\mathrm{lim}} \frac{d}{dx}\left[\frac{1-F(x)}{f(x)}\right]=0$ and $a_K$ and $b_K$ are given by
\begin{align}
    a_K&=\frac{\lambda}{4}\left(\mathrm{ln}\left(K\sqrt{\frac{\pi}{2}}\right)+\frac{1}{2}\mathrm{ln}\left(\mathrm{ln}\left(K\sqrt{\pi}\lambda^{-\frac{1}{4}}\right)\right)\right)^2,\label{a_k}\\
    b_K&=\frac{\lambda}{2}\mathrm{ln}\left(K\sqrt{\pi}\lambda^{-\frac{1}{4}}\right).\label{b_k}
\end{align}
From the limiting throughput distribution theorem in \cite{song2006asymptotic}, $\frac{R_c-\bar{a}_K}{\bar{b}_K}$ also converges uniformly in distribution to a normalized Gumbel RV and the normalizing constants are given by
\begin{align}
    \bar{a}_K&=\frac{1}{N}\mathrm{log}_2\left(1+\frac{Np\alpha^2}{\sigma^2}a_K\right),\label{a_k_bar}\\
    \bar{b}_K&=\frac{1}{N}\mathrm{log}_2\left(\frac{1+\frac{Np\alpha^2}{\sigma^2}\left(a_K+b_K\right)}{1+\frac{Np\alpha^2}{\sigma^2}a_K}\right).\label{b_k_bar}
\end{align}
We further obtain
\begin{align}
    &\mathbb{E}\left(R_c^{SDA}\right)=E_0\bar{b}_K+\bar{a}_K=\frac{E_0}{N}\mathrm{log}_2\left(1+\frac{Np\alpha^2}{\sigma^2}\left(a_K+b_K\right)\right)+\frac{1-E_0}{N}\mathrm{log}_2\left(1+\frac{Np\alpha^2}{\sigma^2}a_K\right).\label{Asymp}
\end{align}
Combing (\ref{a_k}), (\ref{b_k}), and (\ref{Asymp}), we can obtain that the ergodic rate of IoT transmission in SDA scheme scales with $\mathrm{ln}\left(\mathrm{ln} K\right)$. The slope in terms of ergodic sum rate of IoT transmission can be given by $S_{c,SDA}\approx \frac{0.332}{N}$.
\begin{remark}
The ergodic sum rate of IoT transmission for the SA scheme scales with $\mathrm{ln} K$, while for the SDA scheme it scales with $\mathrm{ln}\left(\mathrm{ln} K\right)$. This demonstrates the SA scheme's advantage, since it allows for a higher IoT transmission rate, especially when the number of IoT devices $K$ is large.
\end{remark}

\section{Outage Probability}\label{sec_outage}
Outage probability is an important measure in the design of wireless systems to operate in a fading environment. In this section, we investigate the outage probabilities of the cellular and IoT transmissions in the simultaneous access and selection diversity access schemes.
\subsection{Outage Probability Analysis for Simultaneous Access Scheme}\label{outage_SA}
\subsubsection{Outage Probability of Cellular Transmission}
The outage events of cellular transmission occur when its rate $R_s^{SA}$ is lower than a predefined quality of service rate $\tilde{R}_s$. The outage probability of cellular transmission is given by
\begin{align}
    &P_{out,r}^{SA}=P\left\{R_s^{SA} \leq \tilde{R}_s\right\}.\label{out_original}
\end{align}
By introducing another exponential integral function $\mathrm{E}_1(x)=\int_{x}^{\infty}\frac{e^{-u}}{u}du=-\mathrm{Ei}(-x)$ and using its first-order approximation, we obtain a tight upper bound of $R_s^{SA}$ from (\ref{Rs_original_highSNR}) as
\begin{align}
    R_s^{SA}\leq& \mathrm{log}_2 \left(\frac{p|h_0|^2}{\sigma^2}\right)+ \bigg(-\mathrm{ln}\frac{|h_0|^2}{\alpha^2\sum_{i=1}^K|g_ih_i|^2}+\frac{|h_0|^2}{\alpha^2\sum_{i=1}^K|g_ih_i|^2}-E_0\bigg)\mathrm{log}_2 e\nonumber \\
    =& \frac{|h_0|^2}{\alpha^2\sum_{i=1}^K|g_ih_i|^2}\mathrm{log}_2 e+\mathrm{log}_2\left( \frac{p\alpha^2\sum_{i=1}^K|g_ih_i|^2}{\sigma^2}\right)-E_0\mathrm{log}_2 e.
\end{align}
If the upper bound of $R_s^{SA}$ is lower than $\tilde{R}_s$, $R_s^{SA} \leq \tilde{R}_s $ can also be guaranteed. The closed-form expression of outage probability of cellular transmission can be obtained from the following theorem.
\begin{theorem}
The outage probability of cellular transmission is given by
\begin{align}
    P_{out,r}^{SA}&=F_{\hat{\Lambda}}(2^{a_2})-\frac{ \pi a_32^{a_2}}{n}\sum_{i=1}^{n}o\left(\phi_i\right),\label{out_s_final_SA}
\end{align}
where $a_1=\frac{\mathrm{log}_2 e}{\alpha^2}$, $a_2=\tilde{R}_s-\mathrm{log}_2\left( \frac{p\alpha^2}{\sigma^2}\right)+E_0\mathrm{log}_2 e$, $a_3=\frac{m_{1,\Lambda}^{2m_{1,\Lambda}}}{\Gamma^2(m_{1,\Lambda}){(K\lambda)}^{m_{1,\Lambda}}}$, $m_{1,\Lambda}=m_{2,\Lambda}=\frac{K+\sqrt{K^2+3K}}{3},\Omega_{\Lambda}=K\lambda$, $G\left[\cdot \right]$ is Meijer's G function, $\Gamma(x)$ is Gamma function, and $F_{\hat{\Lambda}}(x)$ and $o(x)$ are given by
\begin{align}
F_{\hat{\Lambda}}(x)&=\frac{1}{\Gamma(m_{1,\Lambda})\Gamma(m_{2,\Lambda})}
\MeijerG*{2}{1}{1}{3}{1}{m_{1,\Lambda},m_{2,\Lambda},0}{\frac{m_{1,\Lambda}m_{2,\Lambda}}{\Omega_{\Lambda}}x},\\
o(x)&=e^{-\frac{x\left(a_2-\mathrm{log}_2 x\right)}{a_1\lambda_0}}x^{m_{1,\Lambda}-1}K_0\left(2\sqrt{\frac{m_{1,\Lambda}^2x}{K\lambda}}\right)\sqrt{1-x^2}.
\end{align}
\begin{IEEEproof}
The details are given in Appendix~\ref{proof of outage}.
\end{IEEEproof}
\end{theorem}

\subsubsection{Outage Probability of IoT Transmission}
For IoT transmission, if the sum rate of IoT devices is less than a target rate $\tilde{R}_c$, we define this event as the outage event. The outage probability of IoT transmission is given by
\begin{align}
    P_{out,c}^{SA}&=P\left\{R_c^{SA} \leq \tilde{R}_c\right\}=P\left\{\frac{1}{N}\mathrm{log}_2\left(1+\frac{Np\alpha^2\sum_{k=1}^K|g_kh_k|^2}{\sigma^2}\right) \leq \tilde{R}_c\right\}\nonumber \\
    &=P\left\{\Lambda \leq \delta\right\}=F_{\Lambda}(\delta), \label{out_original_rc}
\end{align}
where $\delta=\frac{\sigma^2\left(2^{N\tilde{R}_c}-1\right)}{Np\alpha^2}$. Combining (\ref{KG_CDF}) and (\ref{out_original_rc}), we have
\begin{align}
    P_{out,c}^{SA}=\frac{1}{\Gamma^2(m_{1,\Lambda})}
\MeijerG*{2}{1}{1}{3}{1}{m_{1,\Lambda},m_{1,\Lambda},0}{\frac{m_{1,\Lambda}^2}{K\lambda}\delta}.\label{out_c_SA}
\end{align}

Since the above outage probability $P_{out,c}^{SA}$ involves Meijer's G function and thus is hard to analyze, we provide an asymptotic expression for it.
Denoting $\gamma_c=\frac{Np\alpha^2\Lambda}{\sigma^2}$, its PDF is given by $f_{\gamma_c}(x)=\frac{\sigma^2}{Np\alpha^2}f_{\Lambda}(\frac{\sigma^2x}{Np\alpha^2})$. According to~\cite{yang2020outage,wang2003simple}, its PDF can be further expressed as $f_{\gamma_c}(x)=a_4 x^{\iota}+o(x)$, where $a_4$ is a positive constant, $\iota$ quantifies the smoothing order of $f_{\gamma_c}(x)$ at the origin. Thus, the asymptotic PDF of $\gamma_c$ is written as
\begin{align}
    f_{\gamma_c}(x)\simeq \frac{2\left(m_{1,\Lambda}m_{2,\Lambda}\sigma^2\right)^{\frac{m_{1,\Lambda}+m_{2,\Lambda}}{2}}x^{\frac{m_{1,\Lambda}+m_{2,\Lambda}}{2}-1}}{\Gamma(m_{1,\Lambda})\Gamma(m_{2,\Lambda})(Np\alpha^2\Omega_{\Lambda})^{\frac{m_{1,\Lambda}+m_{2,\Lambda}}{2}}},
\end{align}
and the corresponding asymptotic outage expression is given by
\begin{align}
    P_{out,c}^{SA}\simeq \frac{4\left(m_{1,\Lambda}m_{2,\Lambda}\sigma^2\right)^{\frac{m_{1,\Lambda}+m_{2,\Lambda}}{2}}\left(2^{N\tilde{R}_c}-1\right)^{\frac{m_{1,\Lambda}+m_{2,\Lambda}}{2}}}{\Gamma(m_{1,\Lambda})\Gamma(m_{2,\Lambda})(m_{1,\Lambda}+m_{2,\Lambda})(Np\alpha^2\Omega_{\Lambda})^{\frac{m_{1,\Lambda}+m_{2,\Lambda}}{2}}}.\label{out_c_SA_asym}
\end{align}

\subsection{Outage Probability Analysis for Selection Diversity Access Scheme}
\subsubsection{Outage Probability of Cellular Transmission}
Similar to Section \ref{outage_SA}, the outage probability of cellular transmission for the selection diversity access scheme is
\begin{align}
P_{out,r}^{SDA}&=P\left\{\frac{H}{\alpha^2Z}\mathrm{log}_2 e+\mathrm{log}_2\left( \frac{p\alpha^2Z}{\sigma^2}\right)-E_0\mathrm{log}_2 e \leq \tilde{R}_s\right\}.
\end{align}
We can then obtain
\begin{align}
    P_{out,r}^{SDA}=\int_{0}^{2^{a_2}}\int_{0}^{\frac{y\left(a_2-\mathrm{log}_2 y\right)}{a_1}}f_{H}(x) f_{Z}(y)dx dy.\label{outage_SDA}
\end{align}
Substituting the PDF of RVs $H$ and $Z$ into (\ref{outage_SDA}), we have
\begin{align}
    P_{out,r}^{SDA}&=\int_{0}^{2^{a_2}}\left[1-e^{-\frac{y\left(a_2-\mathrm{log}_2 y\right)}{a_1\lambda_0}}\right]f_{Z}(y)dy=F_{Z}(2^{a_2})-B_2\nonumber \\
    &=\left(1-2\sqrt{\frac{2^{a_2}}{\lambda}}K_1\left(2\sqrt{\frac{2^{a_2}}{\lambda}}\right)\right)^K-B_2.\label{out_s_SDA}
\end{align}
Here, $B_2$ is derived as
\begin{align}
    B_2&=\frac{2K}{\lambda}\int_{0}^{2^{a_2}}e^{-\frac{y\left(a_2-\mathrm{log}_2 y\right)}{a_1\lambda_0}}\left(1-2\sqrt{\frac{y}{\lambda}}K_1\left(2\sqrt{\frac{y}{\lambda}}\right)\right)^{K-1}  K_0\left(2\sqrt{\frac{y}{\lambda}}\right)dy =\frac{K \pi 2^{a_2}}{\lambda n}\sum_{i=1}^n q\left(\phi_i\right),
\end{align}
where $q(x)=e^{-\frac{x\left(a_2-\mathrm{log}_2 x\right)}{a_1\lambda_0}}\left(1-2\sqrt{\frac{x}{\lambda}}K_1\left(2\sqrt{\frac{x}{\lambda}}\right)\right)^{K-1} \allowbreak K_0\left(2\sqrt{\frac{x}{\lambda}}\right)\sqrt{1-x^2}$.
\subsubsection{Outage Probability of IoT Transmission}
The outage probability of IoT transmission is given by
\begin{align}
    P_{out,c}^{SDA}&=P\left\{\frac{1}{N}\mathrm{log}_2\left(1+\frac{Np\alpha^2\underset{k}{\max}|g_kh_k|^2}{\sigma^2}\right) \leq \tilde{R}_c\right\}=P\left\{Z \leq \delta\right\}\nonumber \\
    &=\left(1-2\sqrt{\frac{\delta}{\lambda}}K_1\left(2\sqrt{\frac{\delta}{\lambda}}\right)\right)^K.\label{out_c_SDA}
\end{align}

\section{Numerical Results}\label{sec-simulation}
In this section, numerical results are provided to substantiate the performance of the proposed symbiotic radio multiple access systems. The symbol period of the IoT device is $64$ times that of the BS, i.e., $N=64$. The distance between the BS and the receiver is $d_0=162$ m, and the IoT devices are located within a circle with a radius of 10 m centered on the receiver (we denote $d_{k,\text{h}}$ as the distance between the BS and the $k$-th IoT device and $d_{k,\text{g}}$ as the distance between the $k$-th IoT device and the receiver). We assume that the channels are subject to large-scale path loss and small-scale fading, i.e., $h_0=\sqrt{L_0}\eta_0,h_k=\sqrt{L_{k,\text{h}}}\eta_{k,\text{h}},g_k=\sqrt{L_{k,\text{g}}}\eta_{k,\text{g}}$. The path loss model in~\cite{rappaport2010wireless,griffin2009complete} is adopted and the path loss values are given by
\begin{align}
    L_0=\frac{\lambda_c^2G_bG_r}{(4\pi)^2d_0^{\nu_0}},L_{k,\text{h}}=\frac{\lambda_c^2G_bG_i}{(4\pi)^2d_{k,\text{h}}^{\nu_h}},L_{k,\text{g}}=\frac{\lambda_c^2G_rG_i}{(4\pi)^2d_{k,\text{g}}^{\nu_g}},
\end{align}
where $\lambda_c$ is the carrier signal wavelength; $G_b,G_i$ and $G_r$ denote the antenna gain of the BS, IoT device and the receiver, respectively; $\nu_0,\nu_h$ and $\nu_g$ are the path loss exponents for channels $h_0$, $h_k$ and $g_k$. In the simulations, we set $\lambda_c=0.33$ m, corresponding to carrier frequency $f_c=900$ MHz, the antenna gains $G_b=G_i=G_r=0$ dB, the path loss exponents $\nu_0=4,\nu_h=2,\nu_g=2$. The small-scale fadings follow i.i.d. Rayleigh distribution, i.e., $\eta_0\sim {\cal{CN}}\left(0,1\right),\eta_{k,\text{h}}\sim {\cal{CN}}\left(0,0.8\right), \eta_{k,\text{g}}\sim {\cal{CN}}\left(0,1\right)$. Therefore, we have $\lambda_g=L_{k,\text{g}}$, $\lambda_h=0.8L_{k,\text{h}}$, $\lambda_0=L_0$. The target rate for cellular transmission is $\tilde{R}_s=5$ and the target sum rate for IoT transmission is $\tilde{R}_c=0.2$. The noise power is set as $\sigma^2=-110$ dBm. The parameters for Gaussian-Chebyshev quadrature are set as $M_1=M_2=1000$ and $n=128$. The numerical results are obtained by averaging over $10^5$ channel realizations.

\begin{figure}[t]
\begin{subfigure}{0.3\linewidth}
  \centering
  % include first image
  \includegraphics[width=2.19 in]{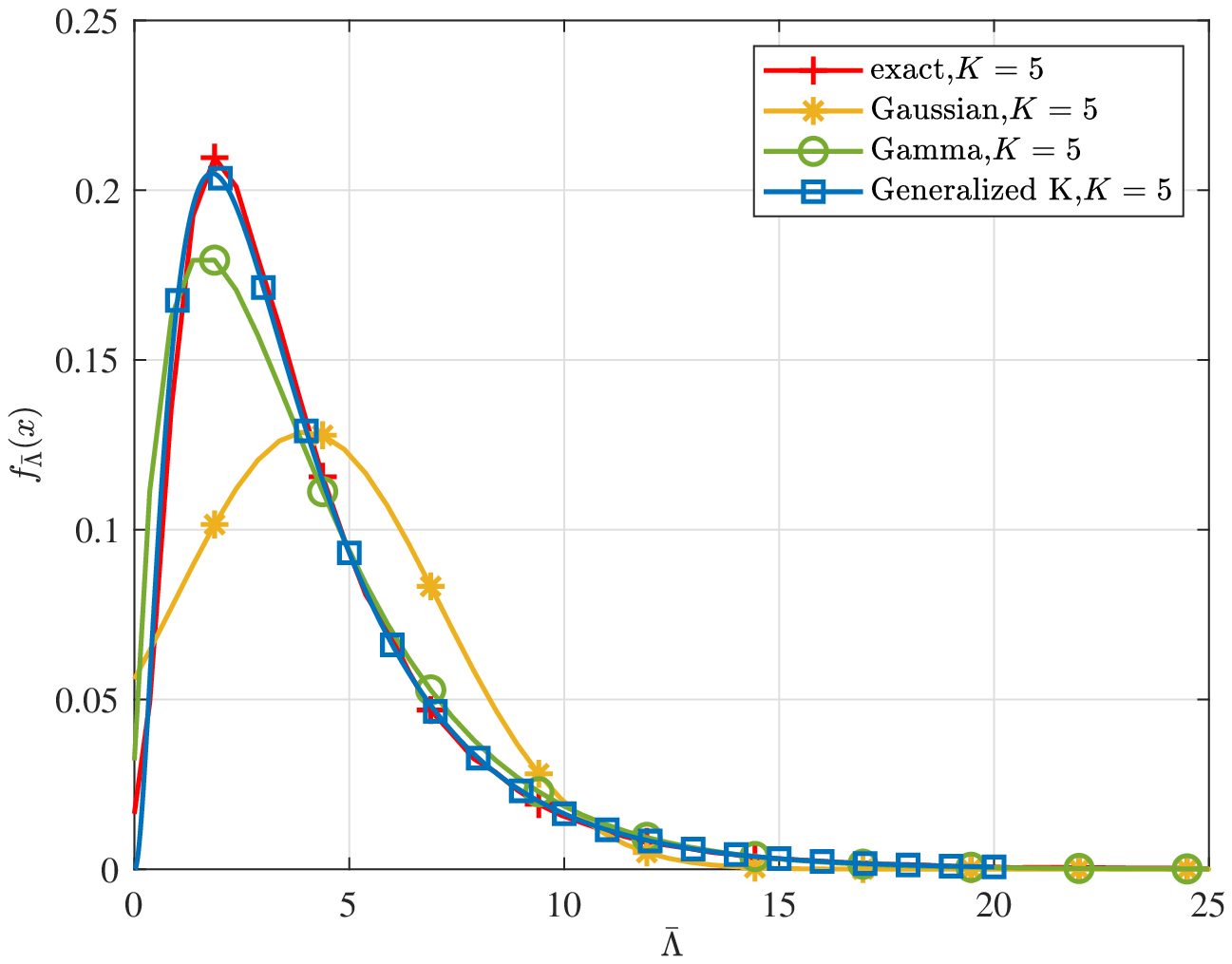}
  \caption{}
  \label{fig:a_K5}
\end{subfigure}
\begin{subfigure}{0.3\linewidth}
  \centering
  % include second image
  \includegraphics[width=2.19 in]{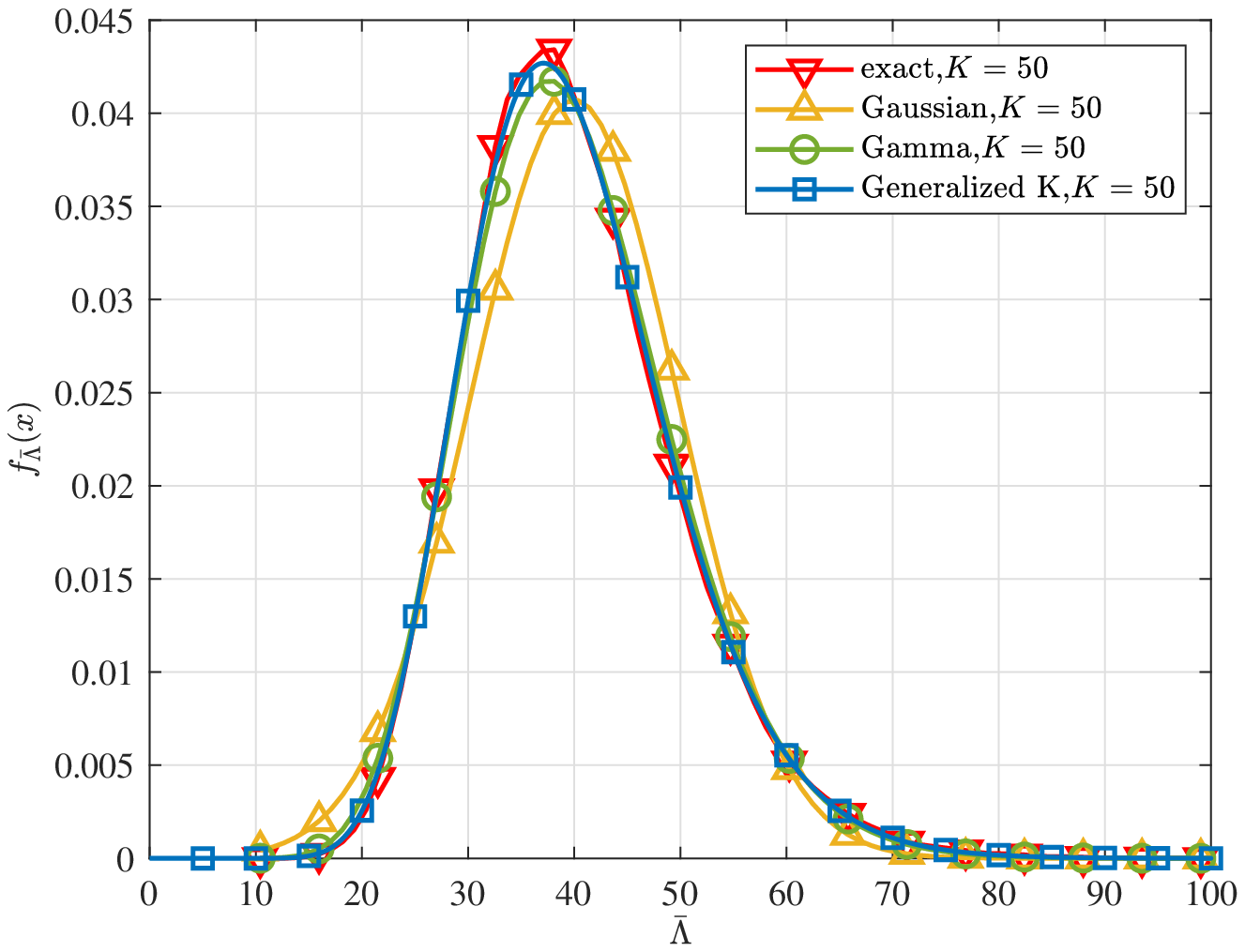}
  \caption{}
  \label{fig:b_K50}
\end{subfigure}
\begin{subfigure}{0.3\linewidth}
  \centering
  % include second image
  \includegraphics[width=2.19 in]{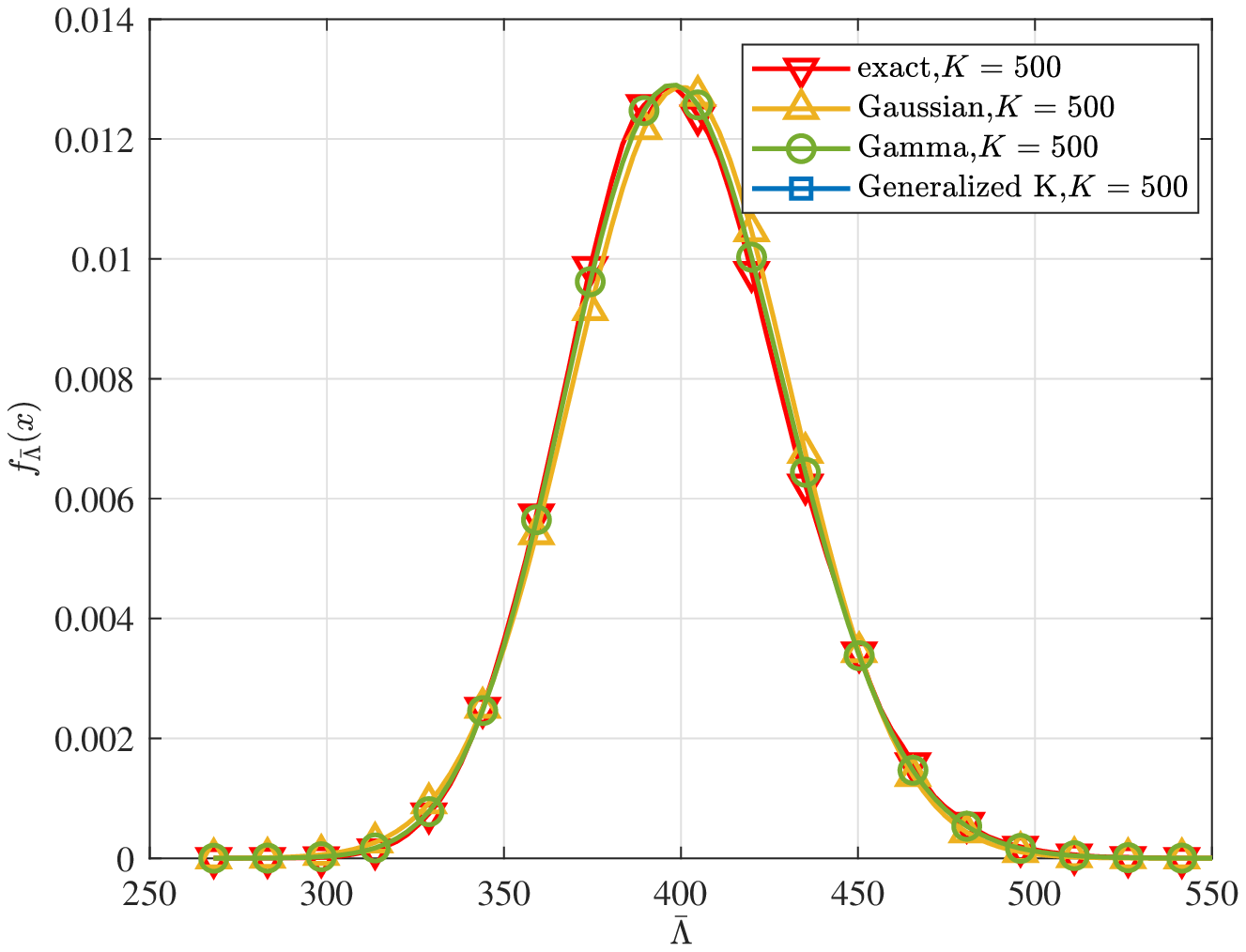}
  \caption{}
  \label{fig:c_K500}
\end{subfigure}
\caption{PDF of $\bar{\Lambda}$ with different $K$ value: (a) $K=5$; (b) $K=50$; (c) $K=500$. }
\label{fig:Fig_lambda}
\end{figure}
Denote $\bar{\Lambda}=\Lambda/10^{-12}$ and Fig.~\ref{fig:Fig_lambda} illustrates the PDF of $\bar{\Lambda}$ when $K$ is varied. The Gamma distribution's parameters $\beta_1$ and $\beta_2$ are given by $\beta_1=\frac{K}{3}$ and $\beta_2=3\lambda$, respectively. When $K$ is small, the most accurate approximation is the generalized-K distribution, as illustrated in Fig.~\ref{fig:a_K5}. Fig.~\ref{fig:b_K50} and Fig.~\ref{fig:c_K500} illustrate how the approximation improves as $K$ increases for all three distributions. For small $K$, the Gaussian approximation does not fit the precise distribution, but for large $K$ it does. In comparison to the other two situations, however, the equation involving Gaussian approximation is simpler to compute. On the basis of the aforementioned considerations, we employ Gaussian approximation for ergodic rate analysis and generalized-K approximation for outage analysis. In what follows, we present the ergodic rate and outage probability results for SA and SDA schemes.

\subsection{Ergodic Rate}
\begin{figure}[t]
\begin{subfigure}{0.5\textwidth}
  \centering
  % include first image
  \includegraphics[width=\linewidth]{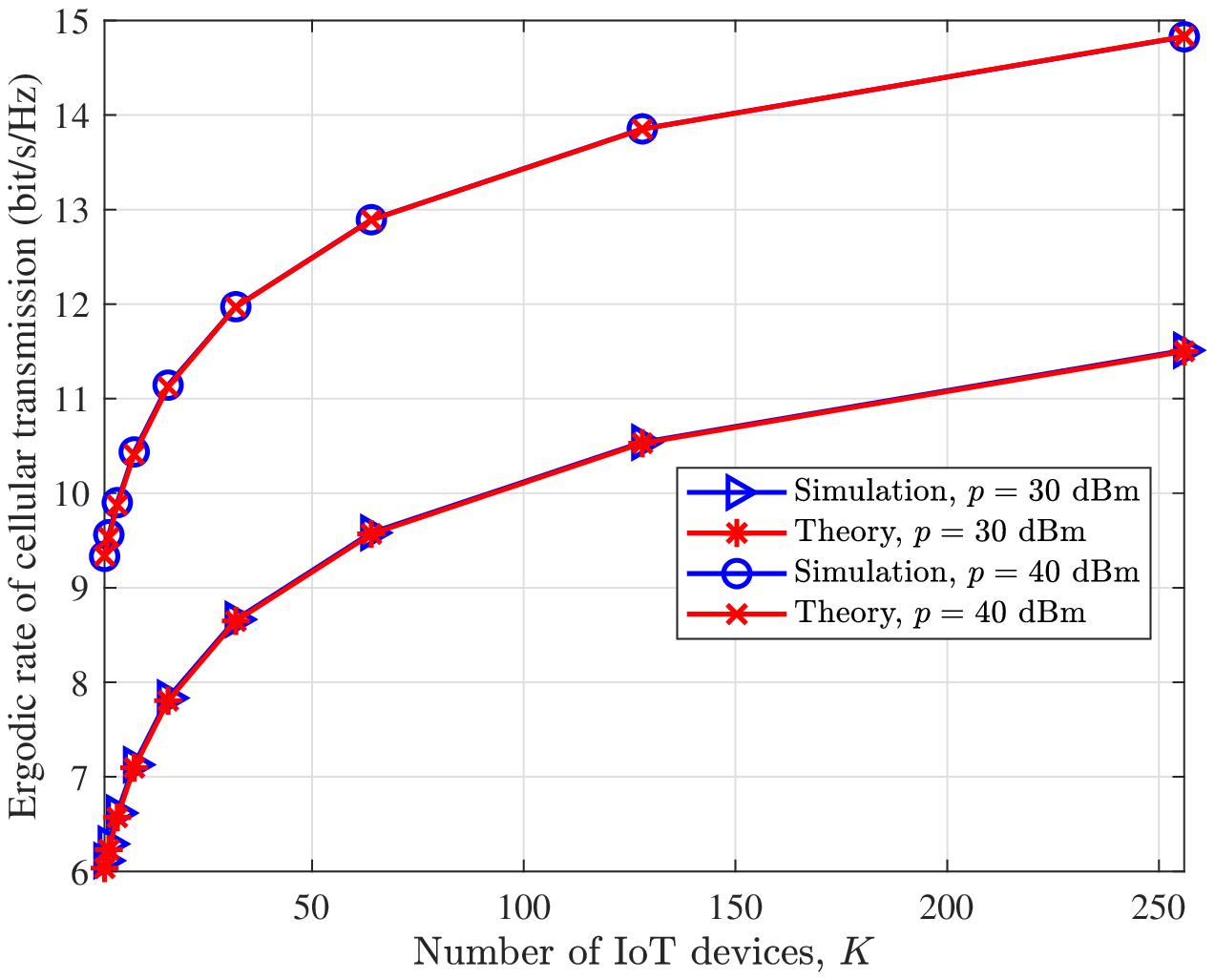}
  \caption{}
  \label{fig:Fig_Gaussian_Rs}
\end{subfigure}
\begin{subfigure}{0.5\textwidth}
  \centering
  % include second image
  \includegraphics[width=\linewidth]{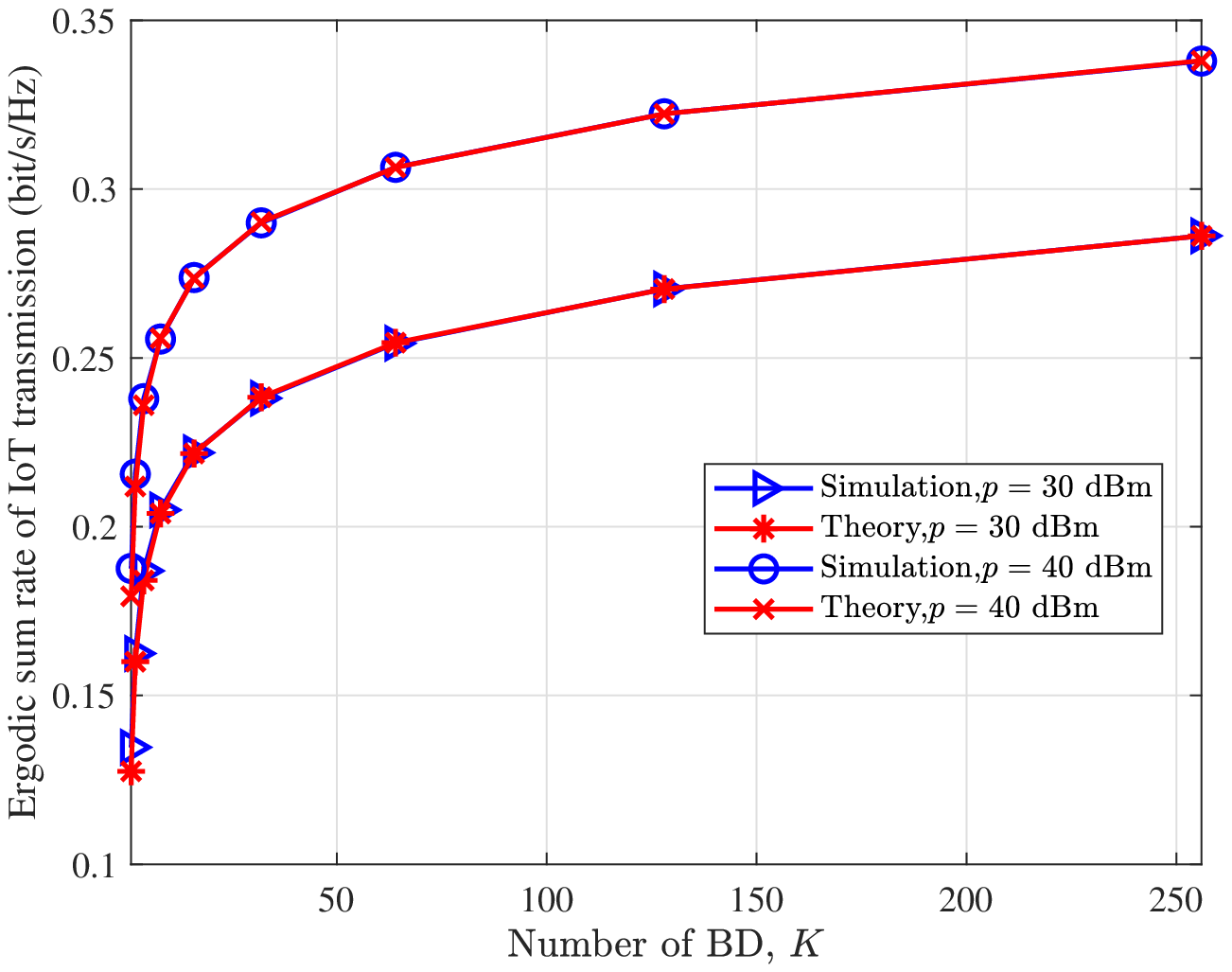}
  \caption{}
  \label{fig:Fig_Gaussian_Rc}
\end{subfigure}
\begin{subfigure}{0.5\textwidth}
  \centering
  % include first image
  \includegraphics[width=\linewidth]{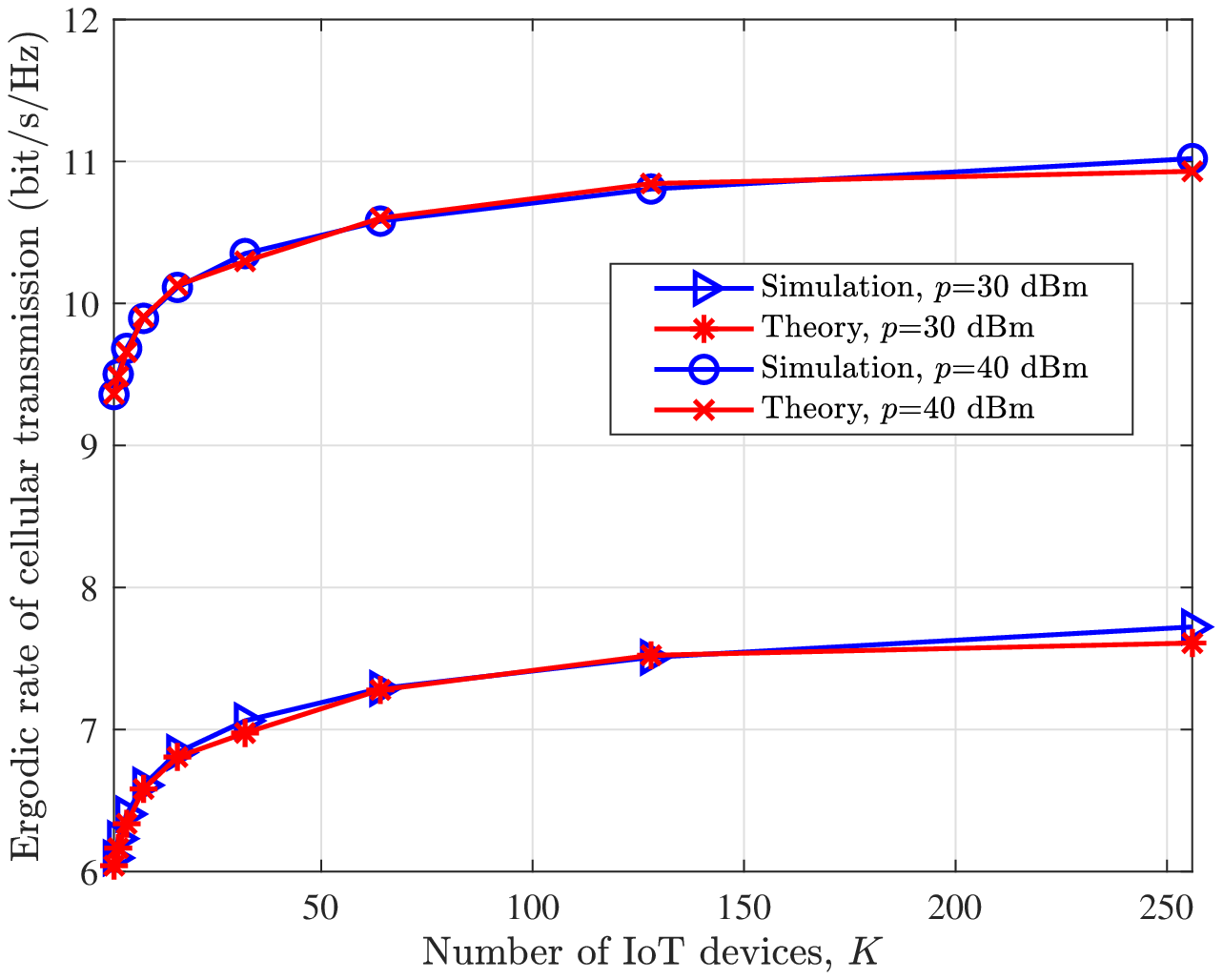}
  \caption{}
  \label{fig:Fig_SDA_Rs}
\end{subfigure}
\begin{subfigure}{0.5\textwidth}
  \centering
  % include second image
  \includegraphics[width=\linewidth]{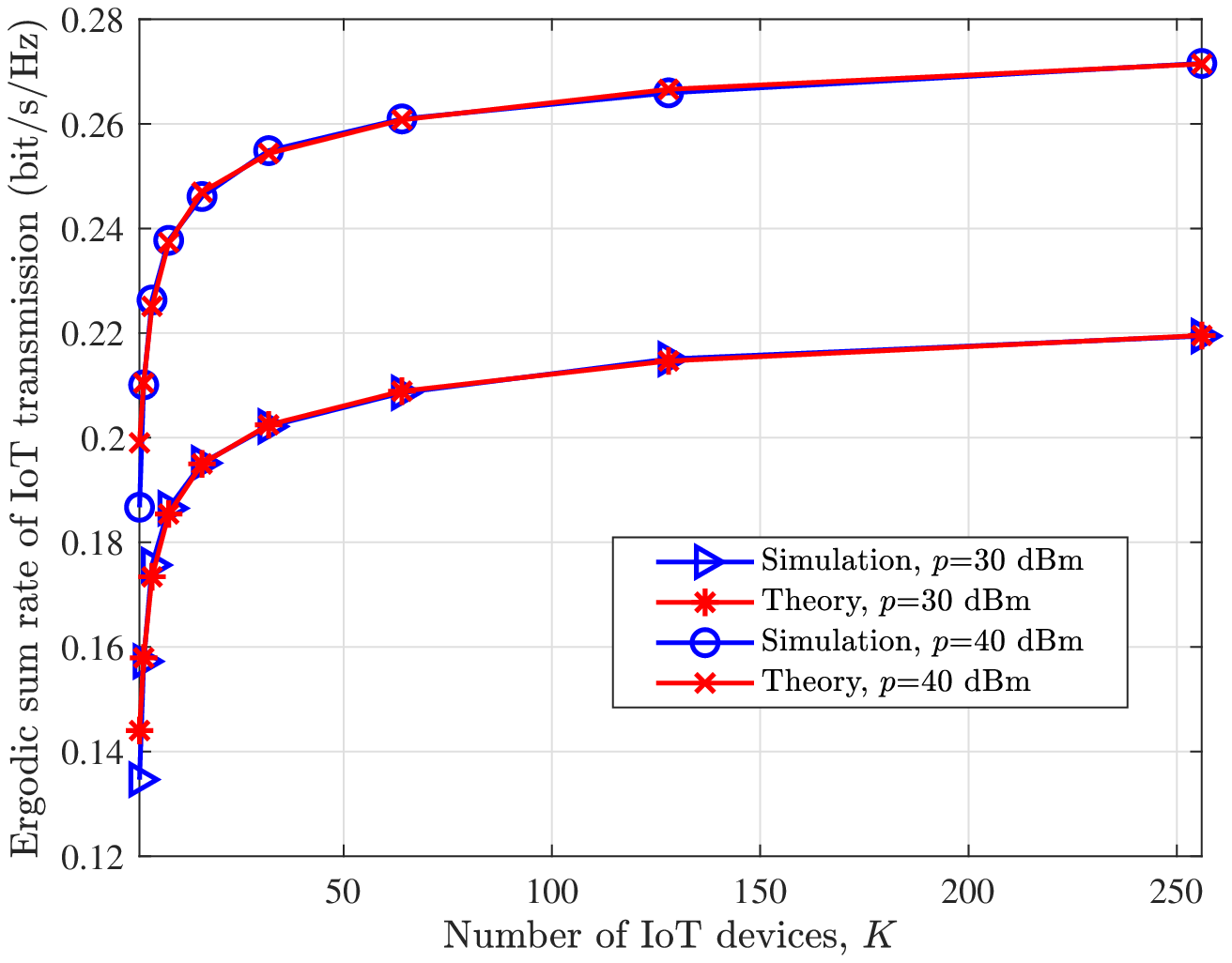}
  \caption{}
  \label{fig:Fig_SDA_Rc}
\end{subfigure}
\caption{ (a) SA scheme: ergodic rate of cellular transmission versus Number of IoT devices; (b) SA scheme: ergodic sum rate of IoT transmission versus Number of IoT devices; (c) SDA scheme: ergodic rate of cellular transmission versus Number of IoT devices ; (b) SDA scheme: ergodic sum rate of IoT transmission versus Number of IoT devices.}
\label{fig:ergodic rate versus K}
\end{figure}

\begin{figure}[!t]
\centering
\includegraphics[width=0.7\columnwidth] {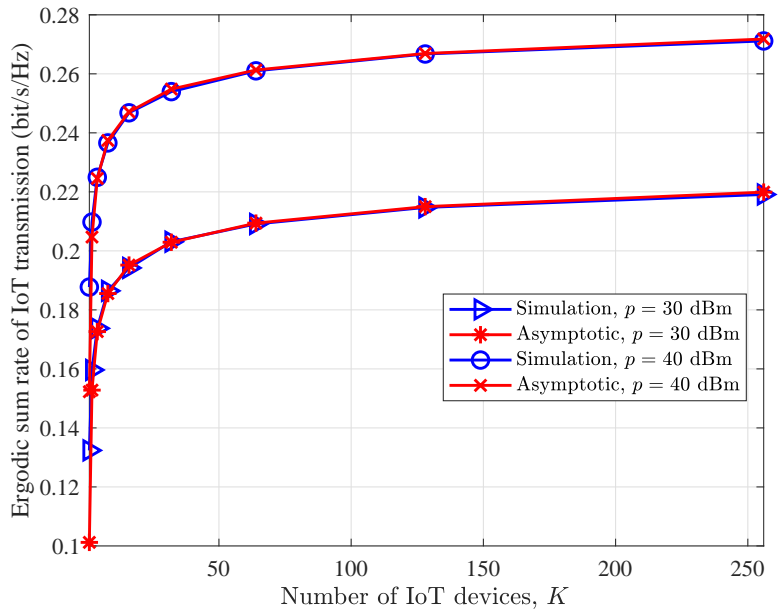}\vspace{-0.3cm}
\caption{Asymptotic ergodic sum rate of IoT transmission versus Number of IoT devices $K$ in SDA scheme.}
\label{fig:Fig_asymptotic}
\vspace{-0.4cm}
\end{figure}

\begin{figure}[t]
\begin{subfigure}{0.5\textwidth}
  \centering
  % include first image
  \includegraphics[width=\linewidth]{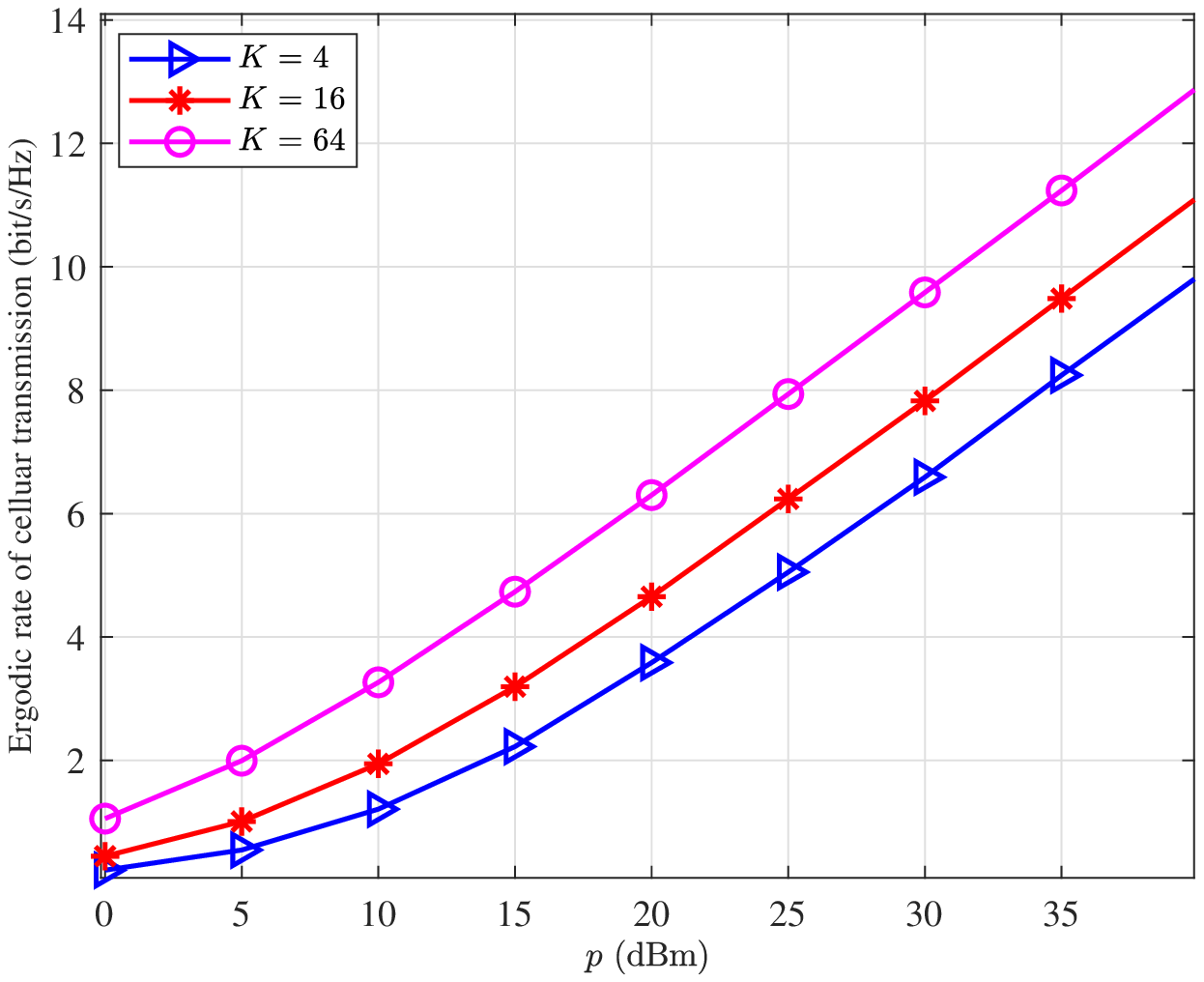}
  \caption{}
  \label{fig:Fig_Rs_vs_snr}
\end{subfigure}
\begin{subfigure}{0.5\textwidth}
  \centering
  % include second image
  \includegraphics[width=\linewidth]{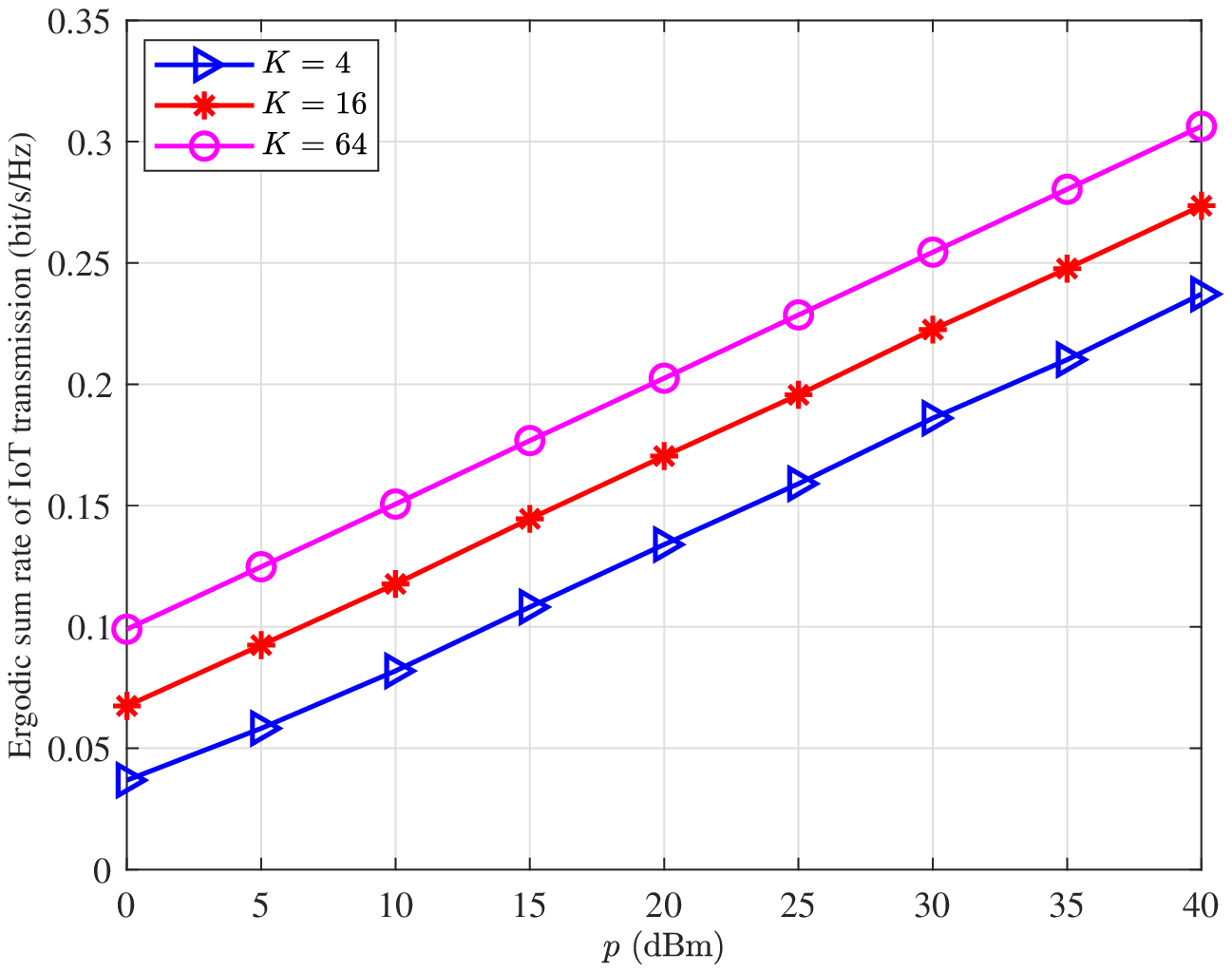}
  \caption{}
  \label{fig:Fig_Rc_vs_snr}
\end{subfigure}
\begin{subfigure}{0.5\textwidth}
  \centering
  % include first image
  \includegraphics[width=\linewidth]{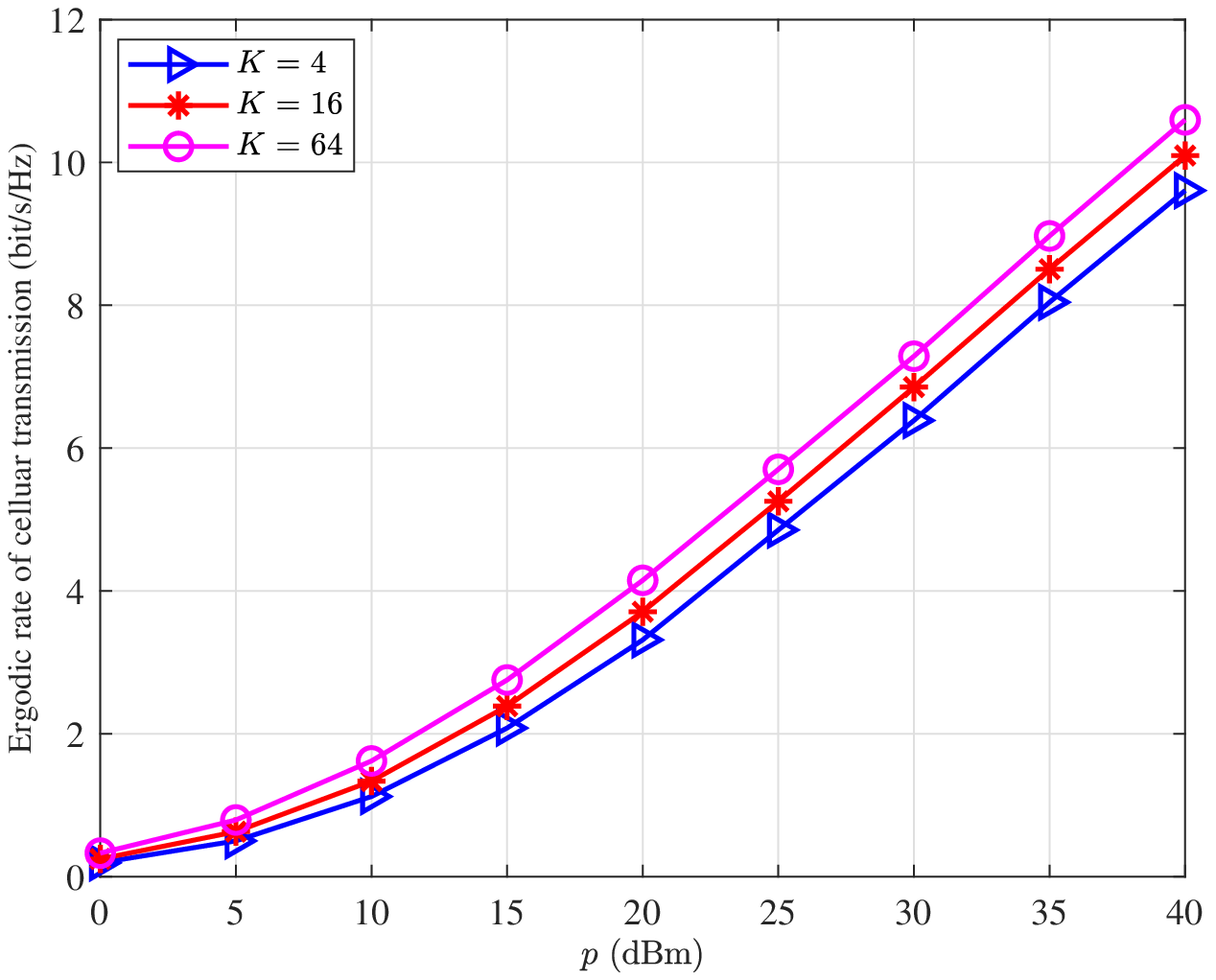}
  \caption{}
  \label{fig:Fig_Rs_vs_snr_SDA}
\end{subfigure}
\begin{subfigure}{0.5\textwidth}
  \centering
  % include second image
  \includegraphics[width=\linewidth]{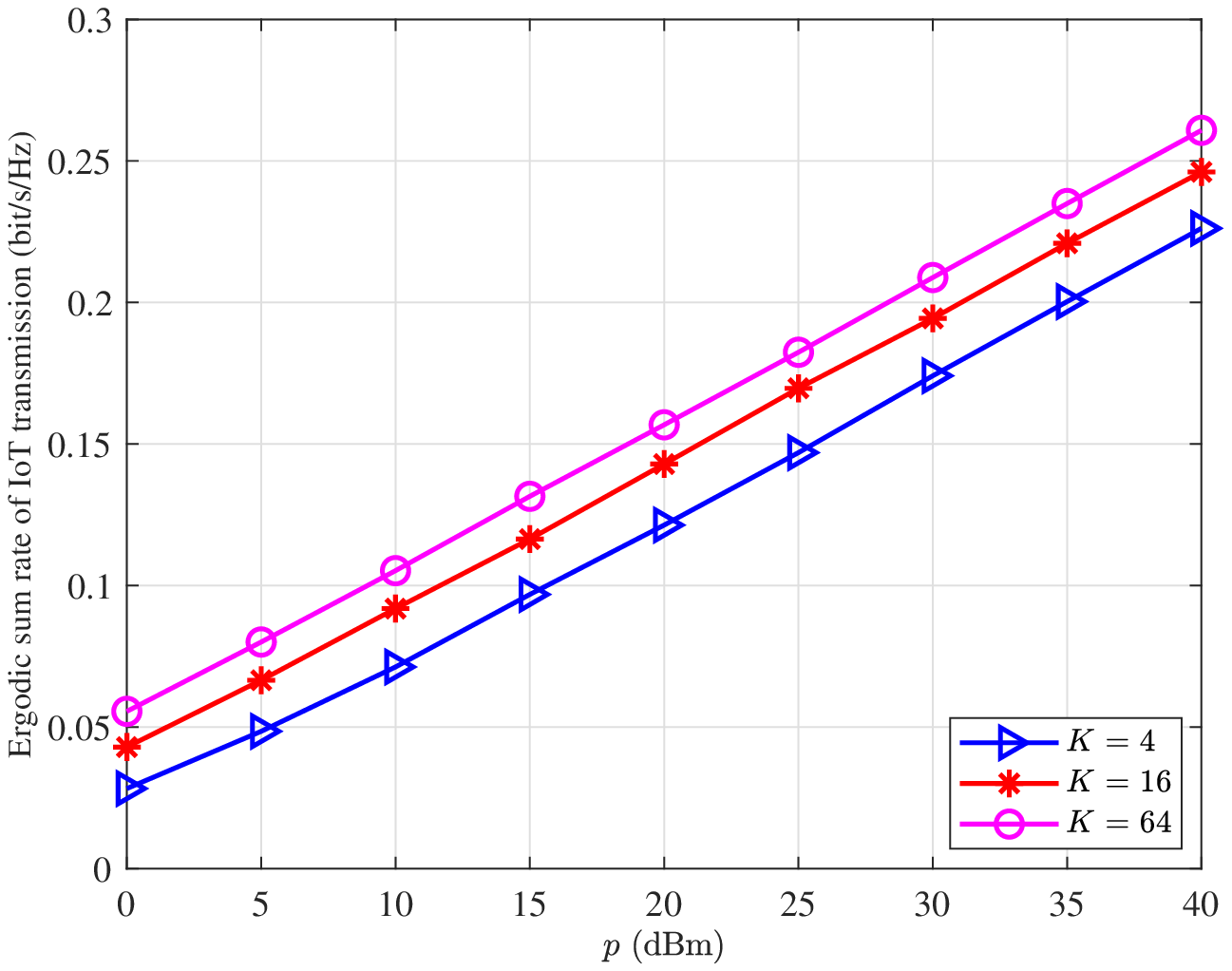}
  \caption{}
  \label{fig:Fig_Rc_vs_snr_SDA}
\end{subfigure}
\caption{ (a) SA scheme: ergodic rate of cellular transmission versus $p$; (b) SA scheme: ergodic sum rate of IoT transmission versus $p$; (c) SDA scheme: ergodic rate of cellular transmission versus $p$; (d) SDA scheme: ergodic sum rate of IoT transmission versus $p$.}
\label{fig:ergodic rate vs snr}
\end{figure}

\begin{figure}[t]
\begin{subfigure}{0.5\textwidth}
  \centering
  % include first image
  \includegraphics[width=\linewidth]{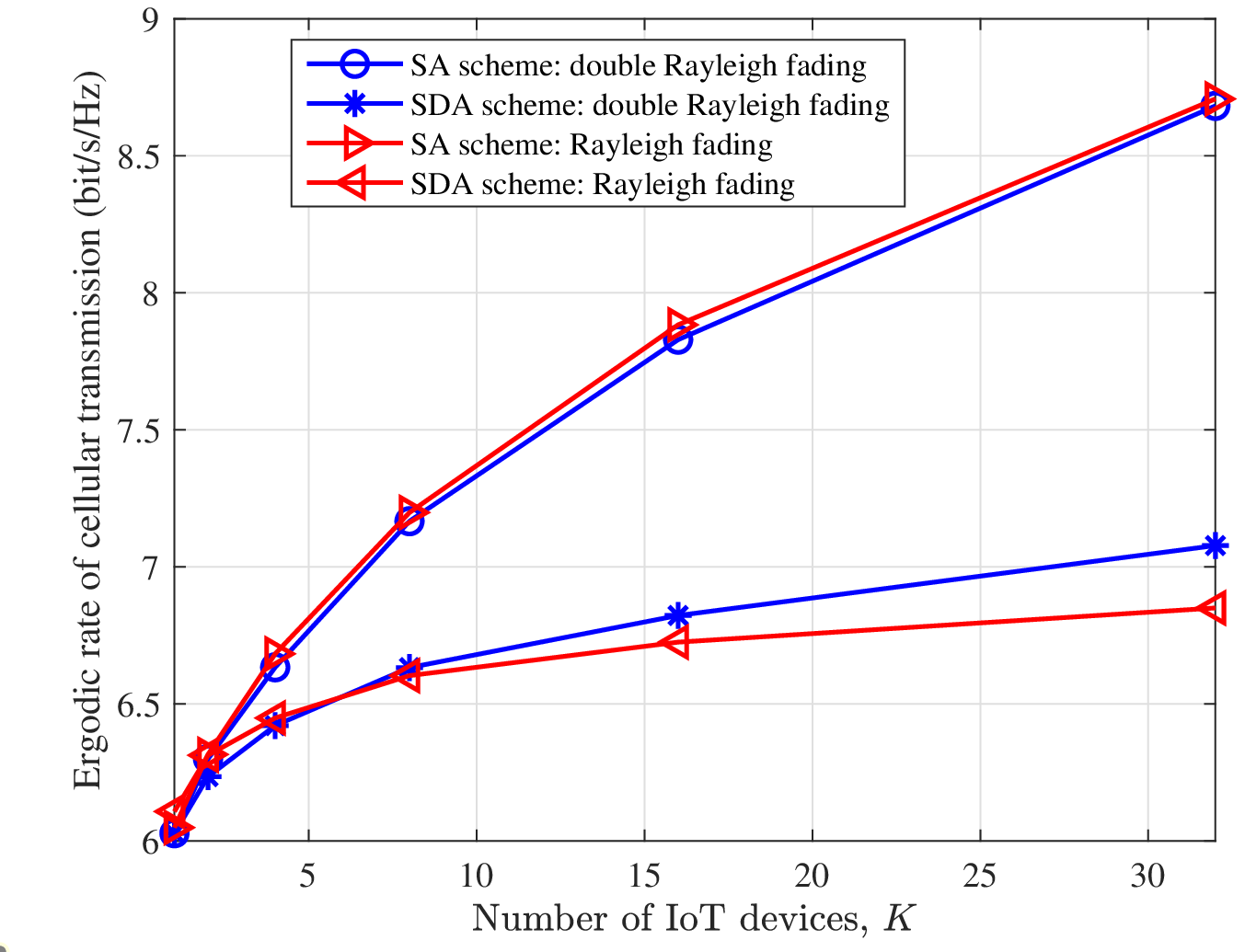}
  \caption{}
  \label{fig:Fig_compare_s}
\end{subfigure}
\begin{subfigure}{0.5\textwidth}
  \centering
  % include second image
  \includegraphics[width=\linewidth]{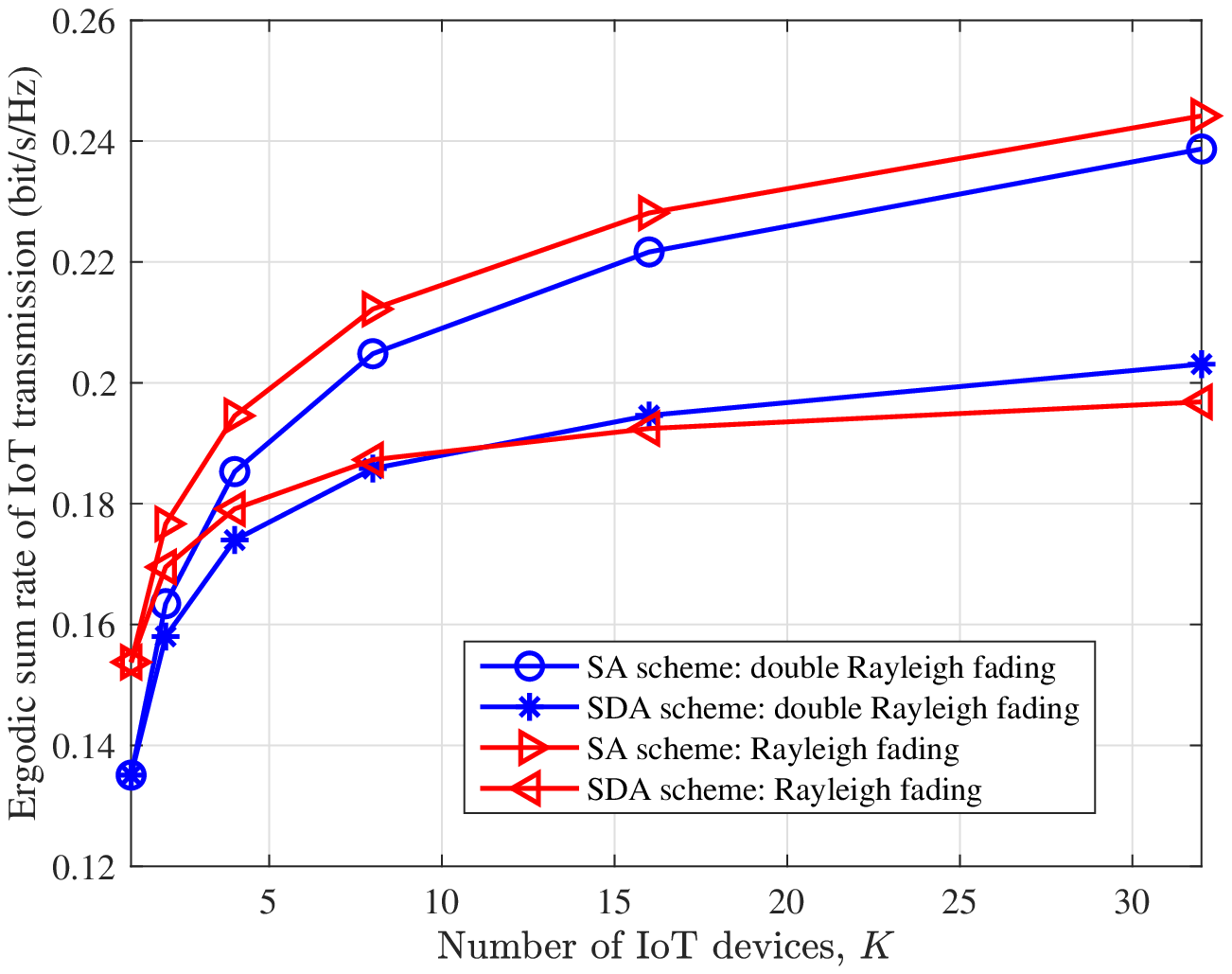}
  \caption{}
  \label{fig:Fig_compare_c}
\end{subfigure}
\caption{ (a) Cellular transmission ergodic rate comparison; (b) IoT transmission ergodic sum rate comparison.}
\label{fig:ergodic rate comparison}
\end{figure}

\begin{figure}[!t]
\centering
\includegraphics[width=0.7\columnwidth] {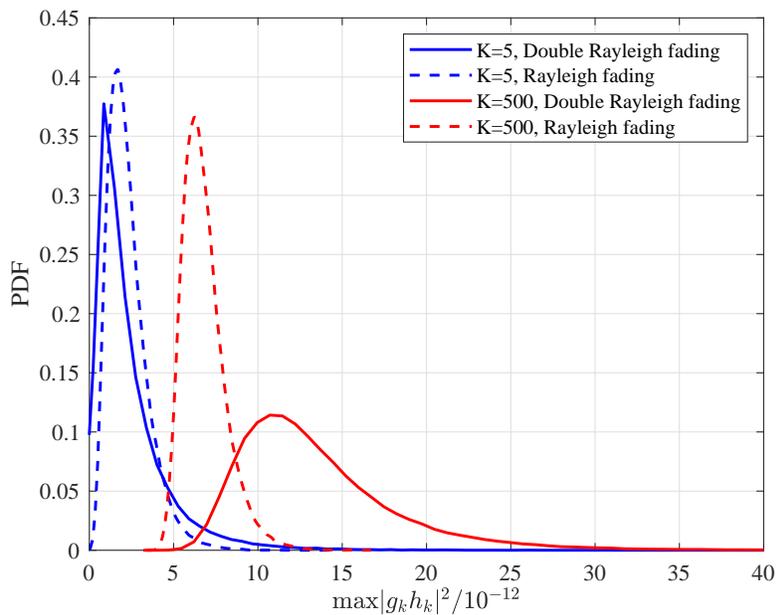}\vspace{-0.3cm}
\caption{PDF of $\underset{k}{\max}|g_kh_k|^2/10^{-12}$ when $K=5$ and $K=500$.}
\label{fig:Fig_compare}
\vspace{-0.4cm}
\end{figure}

%\begin{figure}[!t]
%\centering
%\includegraphics[width=0.7\columnwidth] {multiple_access_Rs_Gaussian_SNR10AND20.eps}\vspace{-0.3cm}
%\caption{Ergodic rate of cellular transmission versus Number of BDs $K$ in SA system.}
%\label{fig:Fig_Gaussian_Rs}
%\vspace{-0.4cm}
%\end{figure}

%\begin{figure}[!t]
%\centering
%\includegraphics[width=0.7\columnwidth] {multiple_access_Rc_Gaussian_SNR10AND20.eps}\vspace{-0.3cm}
%\caption{Ergodic sum rate of BD transmission versus Number of BDs $K$ in SA system.}
%\label{fig:Fig_Gaussian_Rc}
%\vspace{-0.4cm}
%\end{figure}

%\begin{figure}[!t]
%\centering
%\includegraphics[width=0.7\columnwidth] {MBA_outage_s.eps}\vspace{-0.3cm}
%\caption{Outage probability of cellular transmission versus SNR in SA system.}
%\label{fig:Fig_outage_s_SA}
%\vspace{-0.4cm}
%\end{figure}

%\begin{figure}[!t]
%\centering
%\includegraphics[width=0.7\columnwidth] {MBA_outage_c.eps}\vspace{-0.3cm}
%\caption{Outage probability of BD transmission versus SNR in SA system.}
%\label{fig:Fig_outage_c_SA}
%\vspace{-0.4cm}
%\end{figure}
The ergodic rate of cellular transmission and ergodic sum rate of IoT transmission versus the number of IoT devices are depicted in Fig.~\ref{fig:ergodic rate versus K}. (\ref{s_ergodic}) and (\ref{c_ergodic}) are used to create the curves in Fig.~\ref{fig:Fig_Gaussian_Rs} and Fig.~\ref{fig:Fig_Gaussian_Rc}. (\ref{E_Rs_SDA}) and (\ref{E_Rc_SDA}) are used to create the curves in Fig.~\ref{fig:Fig_SDA_Rs} and Fig.~\ref{fig:Fig_SDA_Rc}. It is seen that the theoretical results match the simulation results precisely, which verifies the accuracy of the derived closed-form expressions in Section~\ref{sec_rate}. The performance difference due to the imprecise Gaussian approximation is insignificant. Furthermore, when $K$ increases, both cellular and IoT transmission ergodic rates increase.

In addition, Fig.~\ref{fig:Fig_asymptotic} shows the asymptotic ergodic sum rate of IoT transmission. From Fig.~\ref{fig:Fig_asymptotic}, we observe that the curves of asymptotic results also match well with the simulation results for a large $K$. It is also observed that the asymptotic result is accurate even when $K$ is not that large. Take $K=4$ when $p=30$ dBm as an example, the asymptotic result is $0.1726$ bit/s/Hz while the simulation result is $0.1731$ bit/s/Hz, which is relatively close.

Fig.~\ref{fig:ergodic rate vs snr} illustrates the ergodic rate of cellular transmission and ergodic sum rate of IoT transmission versus BS transmit power $p$. It is seen that a higher transmit power enables a higher ergodic rate, which is consistent with the analysis of ergodic rates in Section~\ref{sec_rate}. %Moreover, the high SNR slope is an important metric to evaluate the influence of channel parameters on the ergodic rate. The slope in high SNR regimes is defined as $S=\underset{p\rightarrow \infty}{\mathrm{lim}}\frac{R_x\left(p\right)}{10\mathrm{log}_{10}p}$, where $x=\left\{s,c\right\}$. Based on the previous analytical results, we have the slope in terms of ergodic rate of cellular transmission as $S_{s,SA}=\frac{\mathrm{log}_2(10)}{10}\approx 0.332$. The slope in terms of ergodic sum rate of IoT transmission is given by
%\begin{align}
    %S_{c,SA}&=\underset{p\rightarrow \infty}{\mathrm{lim}} \frac{\frac{1}{N}\left[\mathrm{ln}\left(1+\frac{K\lambda N\alpha^2p}{\sigma^2}\right)-\frac{3K\lambda^2N^2\alpha^4}{2\left(\frac{\sigma^2}{p}+K\lambda N\alpha^2\right)^2}\right]\mathrm{log}_2 e}{10\mathrm{log}_{10}p}=\frac{\mathrm{log}_2(10)}{10N}\approx \frac{0.332}{N}.
%\end{align}
%For SDA system, the slope in terms of ergodic rate of cellular transmission can be given by $S_{s,SDA}\approx 0.332$. The slope in terms of ergodic rate of IoT transmission can be derived as $S_{c,SDA}\approx \frac{0.332}{N}$. As seen in Fig.~\ref{fig:ergodic rate vs snr}, the slopes of the ergodic rate of cellular transmission are 0.33 and those of the ergodic sum rate of IoT transmission are 0.0052, which is consistent with the analytical results above.
In addition, the performance gain by increasing $K$ in the SDA scheme is not as obvious as the SA scheme.

Fig.~\ref{fig:Fig_compare_s} and Fig.~\ref{fig:Fig_compare_c} depict the ergodic rate comparison between the SA and SDA schemes when $p$ is set to 30 dBm. The blue lines indicate that the backscatter links are subjected to the double Rayleigh fading whereas the red lines indicate that the backscatter links are subjected to typical single Rayleigh fading, i.e., the channels $g_k,\forall k$ are set to $\sqrt{L_{k,\text{g}}}$. We first observe that the ergodic rate of cellular transmission and ergodic sum rate of IoT transmission for SA scheme are larger than those of SDA scheme, which makes sense given that the backscatter link in SA scheme is composed of $K$ individual backscatter links whereas the backscatter link in SDA scheme is the strongest backscatter link. When $K$ equals 16, the ergodic rates of cellular transmission for SA and SDA schemes are 7.82 bit/s/Hz and 6.82 bit/s/Hz, respectively, implying that the performance of SDA scheme can achieve 87.2 percent of that of the SA scheme. This ratio is 87.7 percent in terms of ergodic sum rate of IoT transmission. There exists a trade-off between the computational complexity and performance for the SA and SDA schemes. A significant conclusion from this discovery is that when the number of IoT devices is small, we should use the SDA scheme and allow only the IoT device with the strongest backscatter link to transmit its information while the others remain silent. The reason for this is that its performance can exceed 85 percent of that of the SA scheme while greatly reducing computational complexity and avoiding the interference between IoT devices. When $K$ is large, the SA scheme is preferable since it guarantees a significantly better rate performance and a lower outage probability.

We then compare the SA and SDA schemes with double Rayleigh fading or single Rayleigh fading. For SA scheme, we observe that the ergodic sum rate of IoT transmission with single Rayleigh fading is always larger than that with double Rayleigh fading. The similar observation can be made with the ergodic rate of cellular transmission. The reason for this is that the strength of the backscatter link with single Rayleigh fading $\sum_{k=1}^{K}|h_k|^2$ is always stronger than that with double Rayleigh fading $\sum_{k=1}^{K}|g_kh_k|^2$. The outcomes for the SDA scheme are considerably different. As illustrated in Fig.~\ref{fig:Fig_compare_c}, when $K$ is between 1 and 16, the ergodic sum rate of IoT transmission with single Rayleigh fading is somewhat greater than that with double Rayleigh fading, similar to the SA scheme. When $K$ is larger than 16, however, the ergodic sum rate of IoT transmission with single Rayleigh fading is lower than that with double Rayleigh fading. The main reason is that the multiuser diversity gain depends crucially on the tail of the fading distribution ($\underset{k}{\max}|h_k|^2$ in the Rayleigh fading case and $\underset{k}{\max}|g_kh_k|^2$ in the double Rayleigh fading case)~\cite{tse2005fundamentals}. Fig.~\ref{fig:Fig_compare} shows the PDF of $\underset{k}{\max}|g_kh_k|^2$ normalized by $10^{-12}$ when $K=5$ and $K=500$. When $K=500$, the tail of the distribution with double Rayleigh fading is heavier than that with single Rayleigh fading. The heavier the tail, the more likely there is a IoT device with a very strong backscatter link, and the larger the multiuser diversity gain. Thus, a better ergodic sum rate for IoT transmission can be achieved under double Rayleigh fading. With a stronger backscatter link, the cellular transmission can also be improved. In general, both cellular and IoT transmissions in SDA scheme can benefit from the double fading effect, which conventionally is thought to be detrimental.
%This is because the expectation of $\underset{k}{\max}|h_k|^2$ is smaller than the expectation of $\underset{k}{\max}|g_kh_k|^2$, as shown in Fig.~\ref{fig:Fig_compare}, and the achievable rate is proportional to the expectation of $\underset{k}{\max}|g_kh_k|^2$. This result is consistent with the notion in~\cite{viswanath2002opportunistic} that fading can be benificial because it provides a source of randomness and the more random the better. Thus, both cellular and IoT transmissions in SDA scheme can benefit from the double fading effect, which conventionally is thought to be detrimental to transmission.

\subsection{Outage Probability}

\begin{figure}[t]
\begin{subfigure}{0.5\textwidth}
  \centering
  % include first image
  \includegraphics[width=\linewidth]{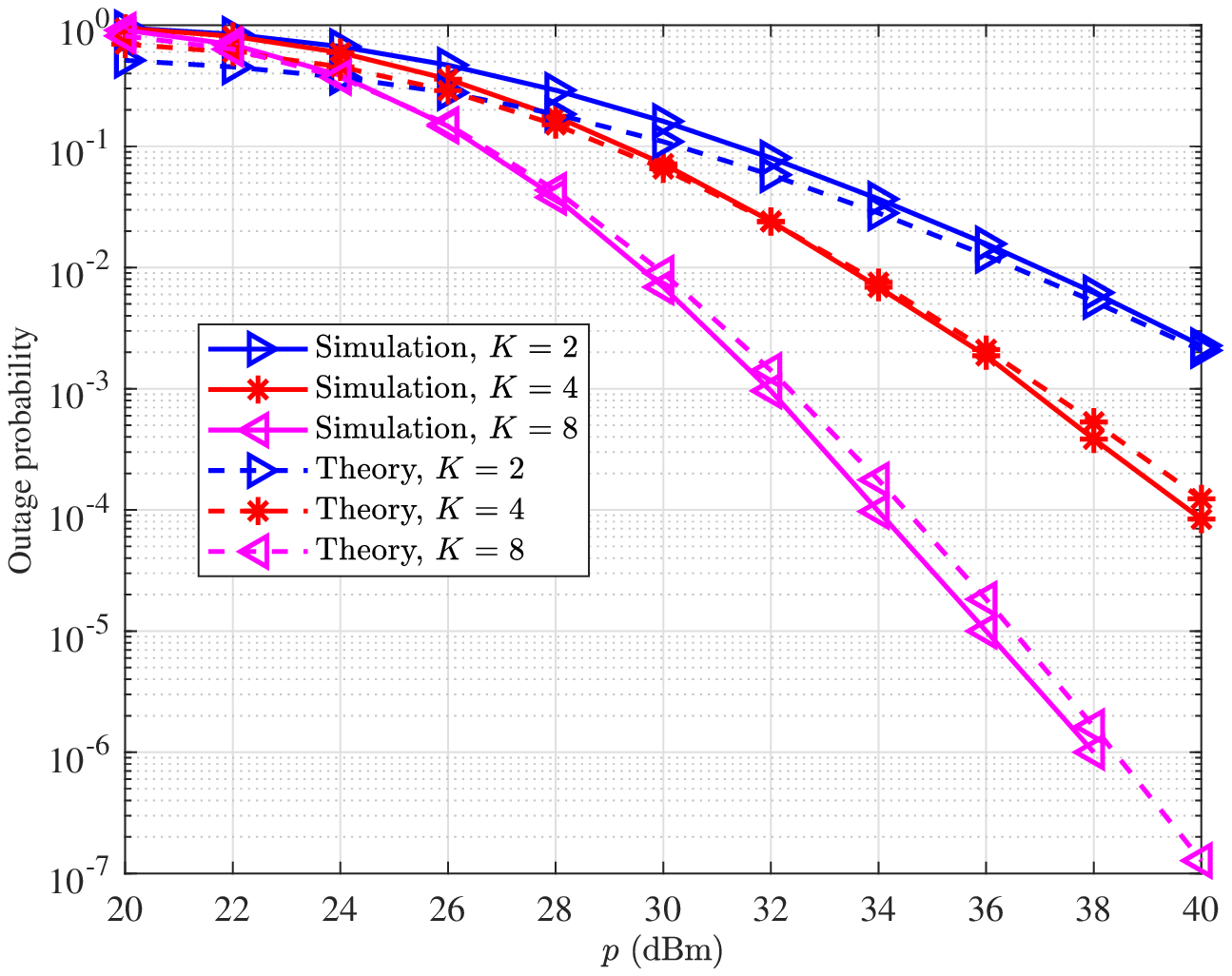}
  \caption{}
  \label{fig:Fig_outage_s_SA}
\end{subfigure}
\begin{subfigure}{0.5\textwidth}
  \centering
  % include second image
  \includegraphics[width=\linewidth]{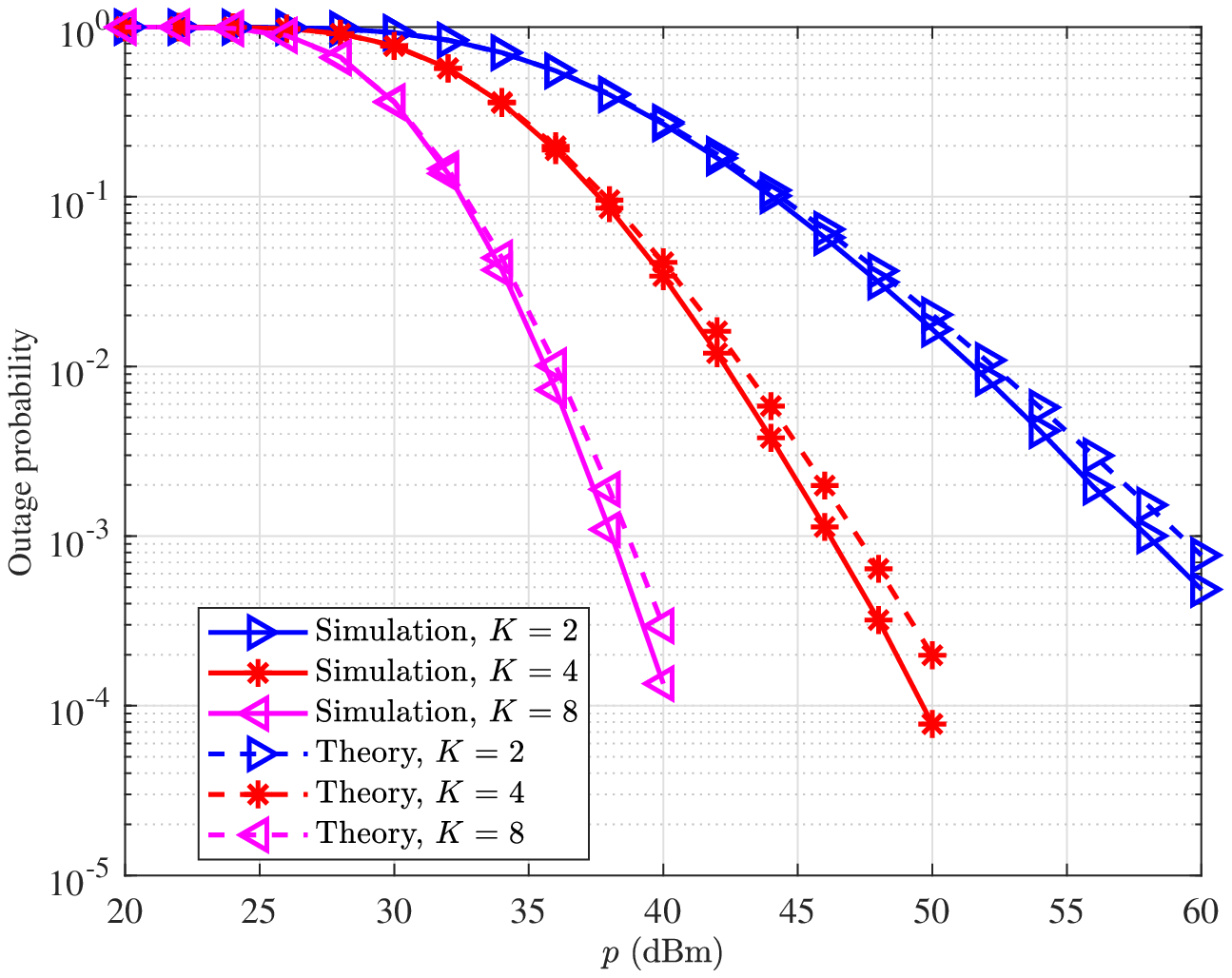}
  \caption{}
  \label{fig:Fig_outage_c_SA}
\end{subfigure}
\begin{subfigure}{0.5\textwidth}
  \centering
  % include first image
  \includegraphics[width=\linewidth]{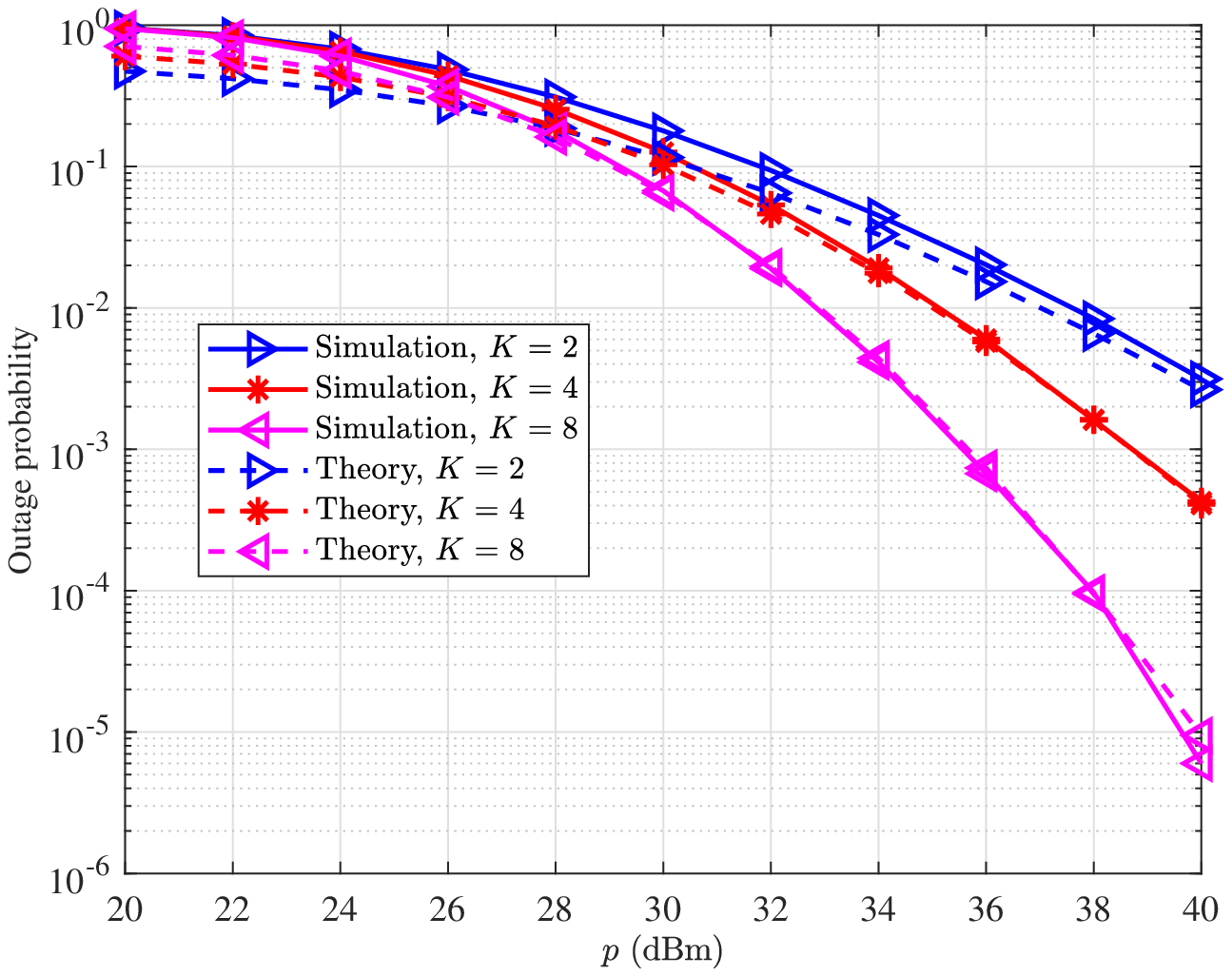}
  \caption{}
  \label{fig:Fig_outage_s_SDA}
\end{subfigure}
\begin{subfigure}{0.5\textwidth}
  \centering
  % include second image
  \includegraphics[width=\linewidth]{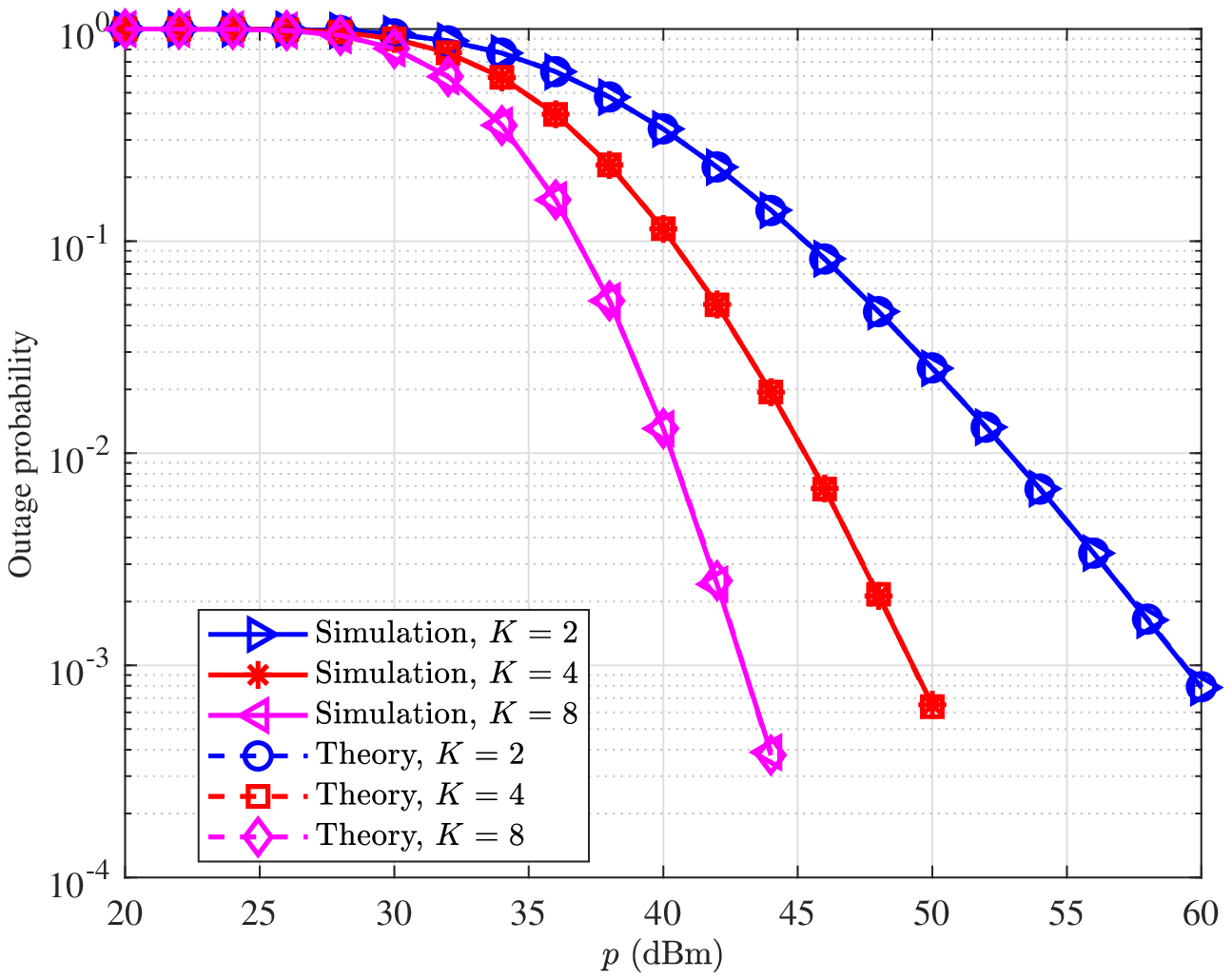}
  \caption{}
  \label{fig:Fig_outage_c_SDA}
\end{subfigure}
\caption{ (a) SA scheme: outage probability of cellular transmission versus $p$; (b) SA scheme: outage probability of IoT transmission versus $p$; (c) SDA scheme: outage probability of cellular transmission versus $p$; (d) SDA scheme: outage probability of IoT transmission versus $p$.}
\label{fig:outage probability}
\end{figure}

\begin{figure}[t]
\begin{subfigure}{0.5\textwidth}
  \centering
  % include first image
  \includegraphics[width=\linewidth]{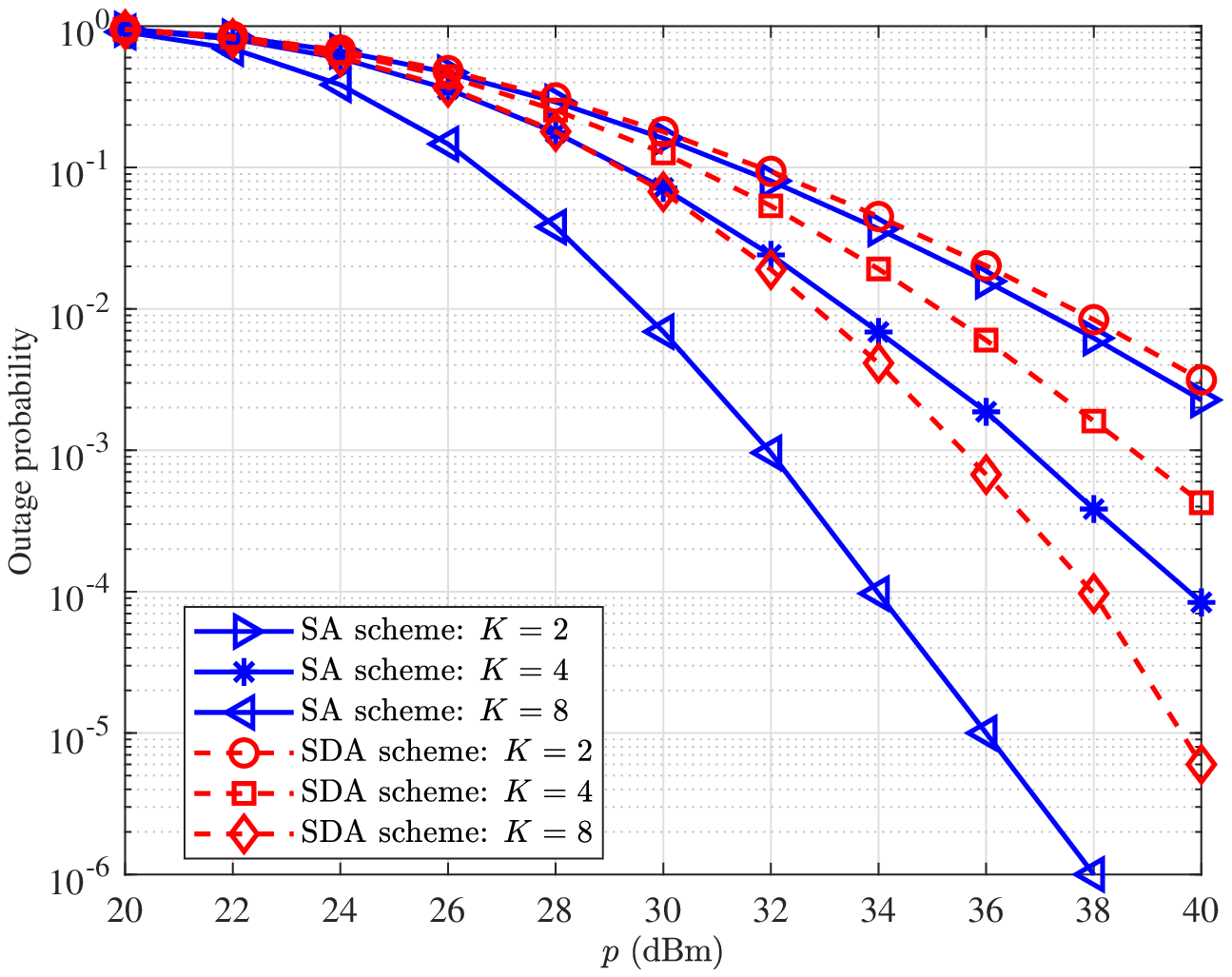}
  \caption{}
  \label{fig:Fig_compare_s_out}
\end{subfigure}
\begin{subfigure}{0.5\textwidth}
  \centering
  % include second image
  \includegraphics[width=\linewidth]{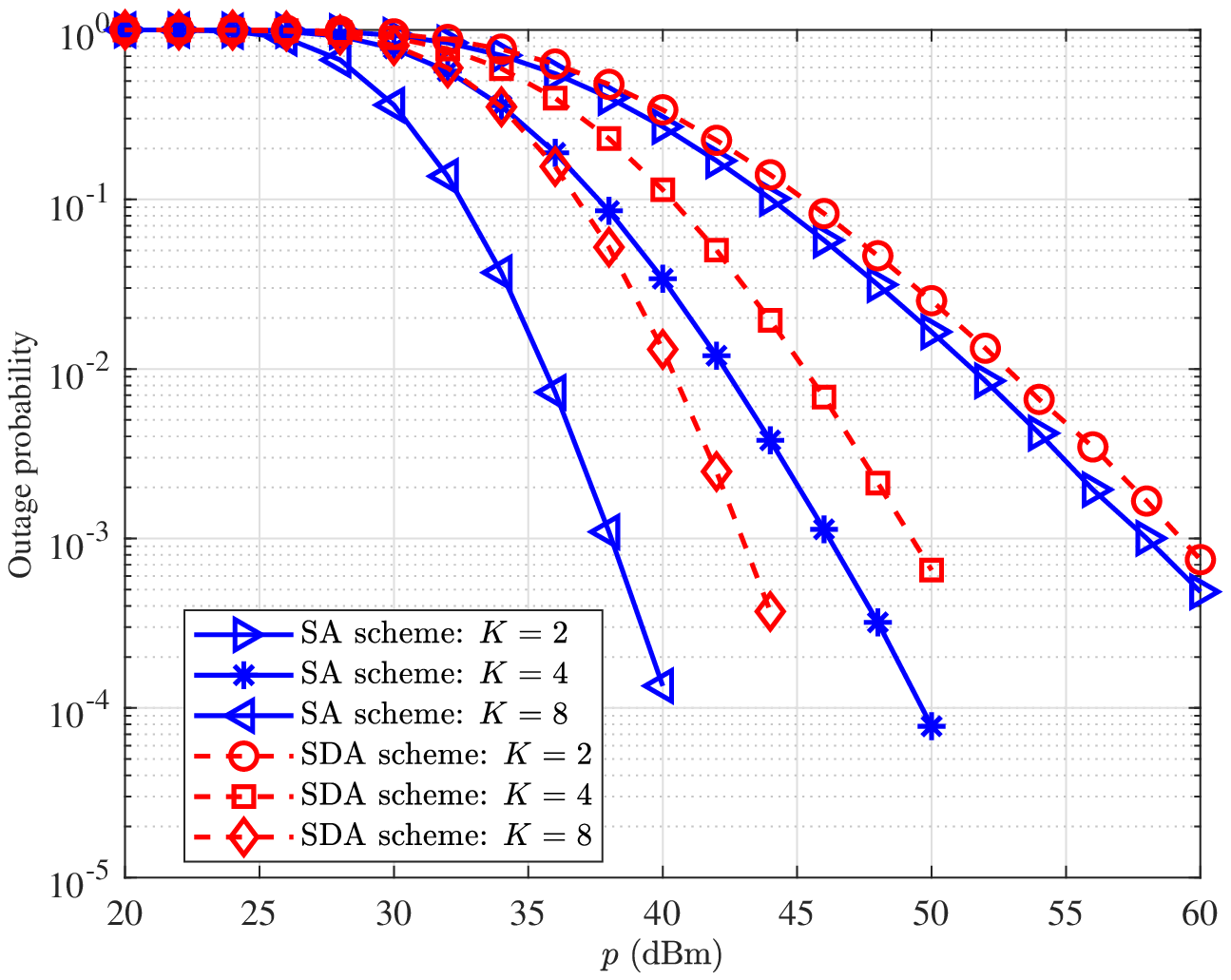}
  \caption{}
  \label{fig:Fig_compare_c_out}
\end{subfigure}
\caption{ (a) Cellular transmission outage probability comparison; (b) IoT transmission outage probability comparison.}
\label{fig:outage probability comparison}
\end{figure}

\begin{figure}[t]
\begin{subfigure}{0.5\textwidth}
  \centering
  % include first image
  \includegraphics[width=\linewidth]{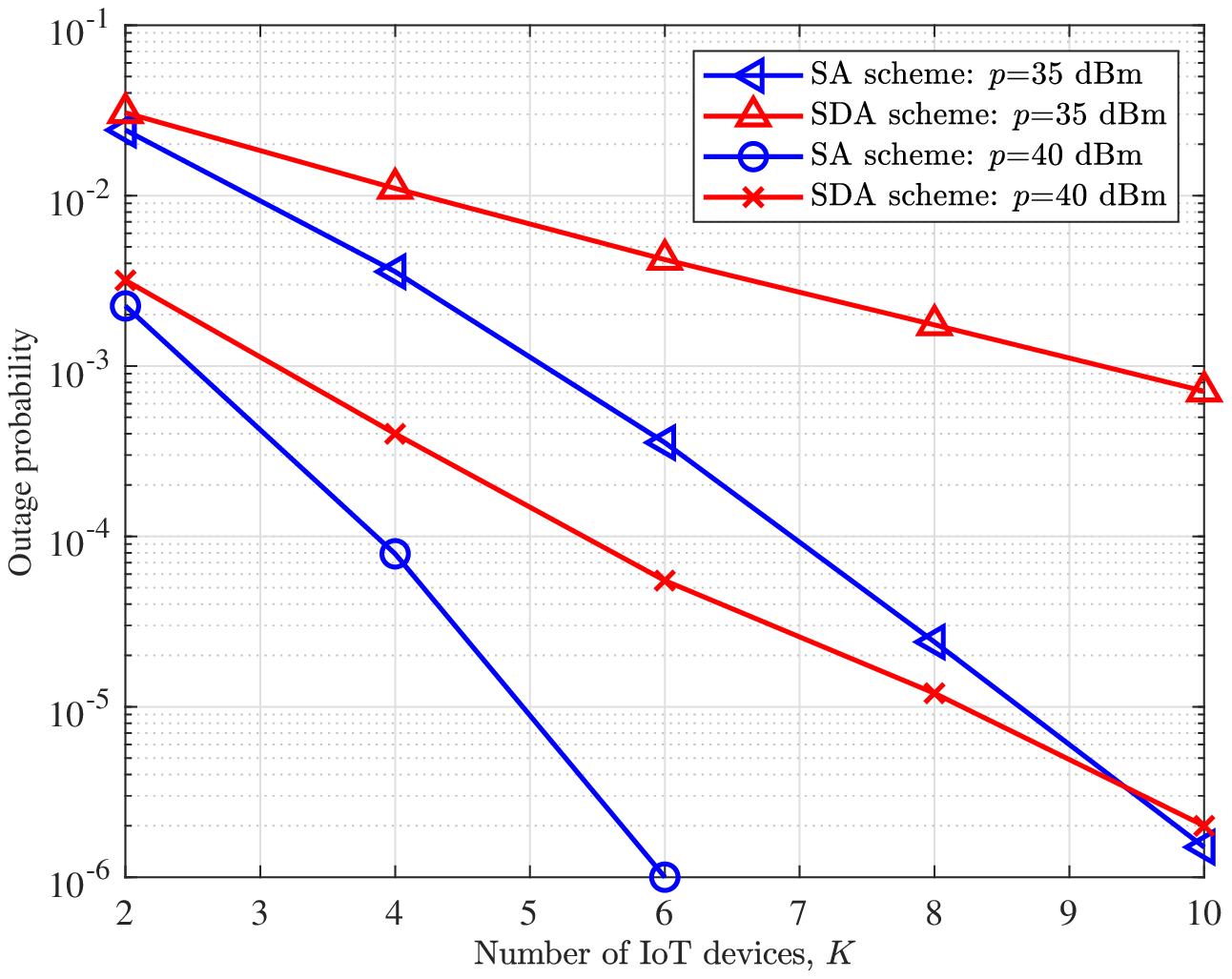}
  \caption{}
  \label{fig:Fig_compare_s_out_K}
\end{subfigure}
\begin{subfigure}{0.5\textwidth}
  \centering
  % include second image
  \includegraphics[width=\linewidth]{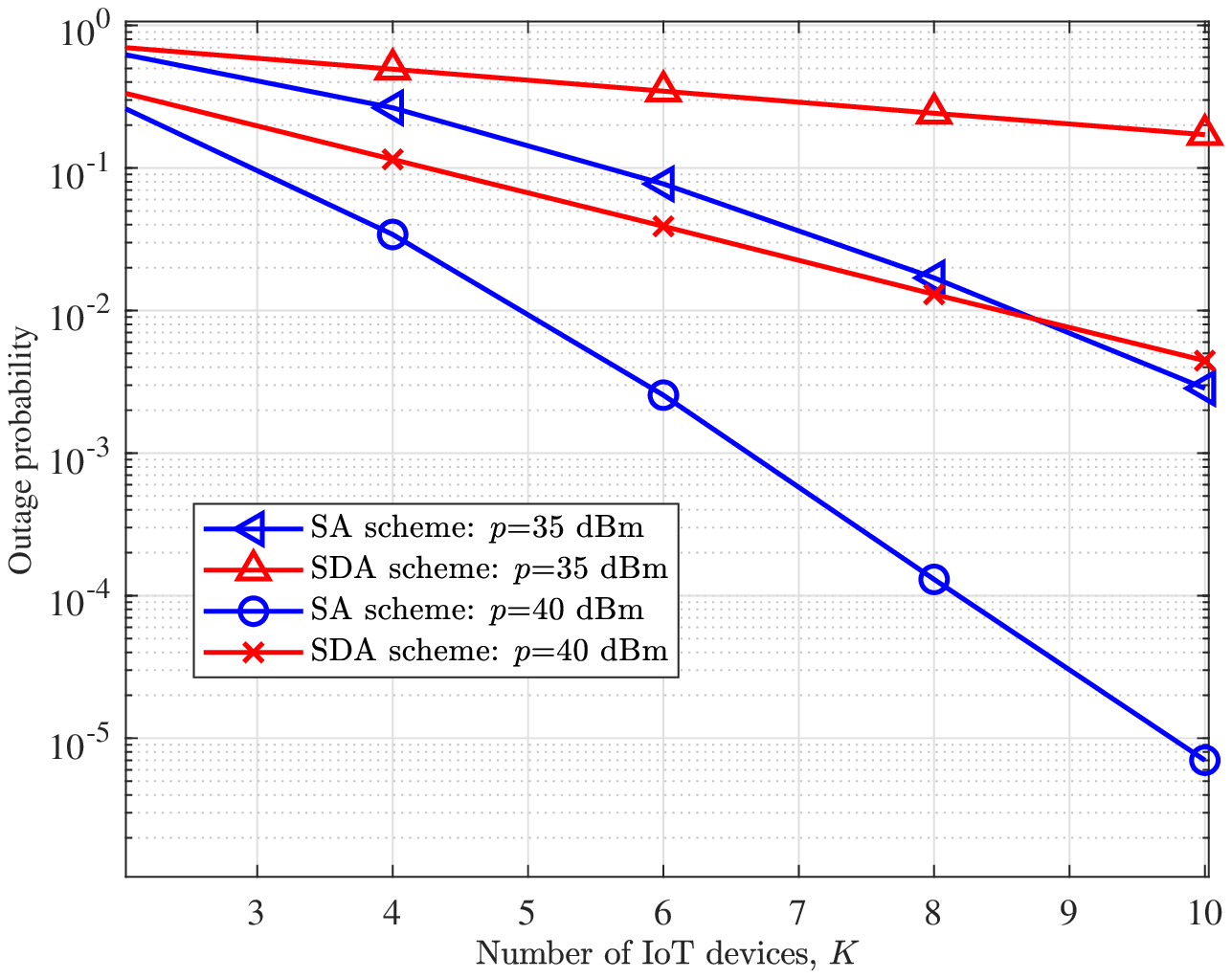}
  \caption{}
  \label{fig:Fig_compare_c_out_K}
\end{subfigure}
\caption{ (a) Outage probability of Cellular transmission versus Number of IoT devices; (b) Outage probability of IoT transmission versus Number of IoT devices.}
\label{fig:outage probability versus K}
\end{figure}

Fig.~\ref{fig:outage probability} shows the outage probabilities of cellular and IoT transmissions in the SA and SDA schemes. The curves in these figures were made using (\ref{out_s_final_SA}), (\ref{out_c_SA}), (\ref{out_s_SDA}), and (\ref{out_c_SDA}). In Fig.~\ref{fig:Fig_outage_s_SA} and Fig.~\ref{fig:Fig_outage_c_SA}, it is seen that when  the outage probability is smaller than $10^{-2}$, there is a minor variation between the theoretical and simulation results, which is caused by the approximation error of $\Lambda$. However, this performance difference is negligible in terms of outage analysis. In Fig.~\ref{fig:Fig_outage_s_SDA} and Fig.~\ref{fig:Fig_outage_c_SDA}, the theoretical results match the simulation results precisely. Additionally, as $K$ increases, the transmit power required to achieve the same outage probability decreases. When transmit power is sufficiently high, the slope is also varied for different values of $K$. The larger $K$ is, the higher the slope is.

Fig.~\ref{fig:outage probability comparison} and Fig.~\ref{fig:outage probability versus K} illustrate the outage probability comparison for cellular and IoT transmissions. It is observed that the outage probabilities of the SA scheme are lower than those of the SDA scheme. The reason is that for a fixed target rate, the rate of the SA scheme is always larger than that of the SDA scheme, leading to a lower outage probability. Moreover, as shown in Fig.~\ref{fig:outage probability comparison}, the outage probability performance of SA scheme is close to that of the SDA scheme when $K$ equals 2. As $K$ increases, the performance gap between the SA and SDA schemes increases. As shown in Fig.~\ref{fig:outage probability versus K}, the outage probabilities for both transmissions decrease with the increase of $K$.

%Fig.~\ref{fig:Fig_outage_s_SDA} and Fig.~\ref{fig:Fig_outage_c_SDA} present the outage probabilities of cellular and BD transmissions in the SDA scheme, respectively. It is seen that the theoretical results match precisely with the simulation results. In addition, with the increase of $K$, the outage probability decreases, which also verifies the benefits of BD selection in SDA scheme. The more BDs can be selected, the better outage performance can be obtained.

%\begin{figure}[!t]
%\centering
%\includegraphics[width=0.7\columnwidth] {MBDA_outage_s.eps}\vspace{-0.3cm}
%\caption{Outage probability of cellular transmission versus SNR in SDA system.}
%\label{fig:Fig_outage_s_SDA}
%\vspace{-0.4cm}
%\end{figure}

%\begin{figure}[!t]
%\centering
%\includegraphics[width=0.7\columnwidth] {MBDA_outage_c.eps}\vspace{-0.3cm}
%\caption{Outage probability of BD transmission versus SNR in SDA system.}
%\label{fig:Fig_outage_c_SDA}
%\vspace{-0.4cm}
%\end{figure}

%\begin{figure}[!t]
%\centering
%\includegraphics[width=0.7\columnwidth] {single_doublefading_compare_s_new.eps}\vspace{-0.3cm}
%\caption{Ergodic rate of cellular transmission comparison.}
%\label{fig:Fig_compare_s}
%\vspace{-0.4cm}
%\end{figure}

%\begin{figure}[!t]
%\centering
%\includegraphics[width=0.7\columnwidth] {single_doublefading_compare_c_new.eps}\vspace{-0.3cm}
%\caption{Ergodic sum rate of BD transmission comparison.}
%\label{fig:Fig_compare_c}
%\vspace{-0.4cm}
%\end{figure}

%by comparing (\ref{Rs_original_highSNR}), (\ref{c_rate}), (\ref{}) and (\ref{})

\section{Conclusions}\label{sec-conclusion}
 In this paper, we have proposed two multiple access schemes, the SA and SDA schemes, for symbiotic radio systems to support massive IoT connections using the cellular networks. We have analyzed the ergodic rates and outage probabilities for both cellular and IoT transmissions and derived closed-form expressions for them. Numerical results have been provided to verify the theoretical analysis and compare the performance of two multiple access schemes. When the number of IoT devices is small, the SDA scheme is preferable since it can significantly reduce computational complexity and avoid interference among IoT devices. On the other hand, when the number of IoT devices is large, the SA scheme is better since it guarantees a significantly higher ergodic rate and lower outage probability performance. In addition, both cellular and IoT transmissions in SDA scheme can
benefit from the double fading effect since a larger multiuser diversity gain can be guaranteed due to a heavier tail of fading distribution.

%\section*{Acknowledgement}
\appendices
\section{Proof of Proposition 1} \label{proof of gamma_s for high SNR}

Denote $T=h_0+\sum_{k=1}^K\alpha_kg_kh_kc_k$. The real part and imaginary part of $T$ are distributed as $\mathrm{Re}\left\{T\right\}\sim {\cal{N}}(\mathrm{Re}(h_0),t)$ and $\mathrm{Im}\left\{T\right\}\sim {\cal{N}}(\mathrm{Im}(h_0),t)$, where $t=\frac{\sum_{k=1}^K|\alpha_kg_kh_k|^2}{2}$. Thus, the RV $x=\frac{|T|^2}{t}$ follows the noncentral chi-square distribution with degree of freedom of $2$ and noncentrality parameter of $\lambda=\frac{|h_0|^2}{t}$. The probability density function (PDF) of $x$ is given by
\begin{align}
    f_X(x)=\frac{1}{2}e^{-\frac{1}{2}\left(x+\lambda\right)}I_0\left(\sqrt{\lambda x}\right),\tag{A1}
\end{align}
where $I_0(\cdot)$ is the modified Bessel function of the first kind, given by $I_0(x)=\sum_{m=0}^{\infty}\frac{1}{m!\Gamma(m+1)}\left(\frac{x}{2}\right)^{2m}$.
The RV $\gamma_s$ can be given by $\gamma_s=\frac{p|T|^2}{\sigma^2}=\frac{pt}{\sigma^2}x$. The PDF of $\gamma_s$ is $f_{\gamma_s}(v)=\frac{\sigma^2}{pt}f_X(\frac{\sigma^2v}{pt})$. Therefore, the PDF of $\gamma_s$ is given by
 \begin{align}
   f_{\gamma_s}(v)=\frac{\sigma^2}{2pt}e^{-\frac{1}{2}\left(\frac{\sigma^2v}{pt}+\frac{|h_0|^2}{t}\right)}I_0\left(\sqrt{\frac{|h_0|^2\sigma^2v}{pt^2}}\right)\triangleq \frac{\sigma^2}{2pt}e^{-\mu-s^2}I_0\left(2s\sqrt{\mu}\right),  \tag{A2}
 \end{align}
where $\mu=\frac{\sigma^2v}{2pt}$ and $s^2=\frac{|h_0|^2}{2t}$.

We have
\begin{align}
    \int_0^{\infty}\mathrm{ln}\mu\left(\frac{\mu}{s^2}\right)^{\frac{m-1}{2}}e^{-\mu-s^2}I_{m-1}\left(2s\sqrt{\mu}\right)d\mu=g_m(s^2),\tag{A3}
\end{align}
where
\begin{align}g_m(x)&=\mathrm{ln}(x)-\mathrm{Ei}(x)+\sum_{j=1}^{m-1}(-1)^j\left[e^{-x}(j-1)!-\frac{(m-1)!}{j(m-1-j)!}\right]\left(\frac{1}{x}\right)^j.\tag{A4}
\end{align}
Here, $\mathrm{Ei}(x)=\int_{-\infty}^{x}\frac{e^u}{u}du$ is the exponential integral. For $m=1$, we have
\begin{align}
    \int_0^{\infty}\mathrm{ln}\mu e^{-\mu-s^2}I_0\left(2s\sqrt{\mu}\right)d\mu=g_1(s^2)=\mathrm{ln}(s^2)-\mathrm{Ei}(s^2).\tag{A5}
\end{align}
For high SNR case, the rate can be given by a closed form, written as
\begin{align}
    R_s^{SA}&\approx \mathbb{E}_{\left\{c_i\right\}_{i=1}^K}\left[\mathrm{log}_2\left(\gamma_s\right)\right]=\int_0^{\infty}\mathrm{log}_2 vf_V(v)dv =\mathrm{log}_2 e\int_0^{\infty}\mathrm{ln}\left(\frac{2pt\mu}{\sigma^2}\right) e^{-\mu-s^2}I_0\left(2s\sqrt{\mu}\right)d\mu \nonumber \\
    &=\mathrm{log}_2 e\mathrm{ln}\left(\frac{2pt\mu}{\sigma^2}\right)+\left[\mathrm{ln}(s^2)-\mathrm{Ei}(s^2)\right]\mathrm{log}_2 e \nonumber \\
    &=\mathrm{log}_2 \left(\frac{p|h_0|^2}{\sigma^2}\right)-\mathrm{Ei}\left(-\frac{|h_0|^2}{\sum_{k=1}^K|\alpha_kg_kh_k|^2}\right)\mathrm{log}_2 e.\tag{A6}
\end{align}

\section{Proof of Theorem 1}\label{ergodic_rate_s_SA}
Denote $H=|h_0|^2$ and $\Psi=\frac{H}{\Lambda}$. The first item in (\ref{eq_Eh}) can be calculated as
\begin{align}
    \mathbb{E}_{\boldsymbol{h_0}}\left[\mathrm{log}_2 \left(\frac{p|h_0|^2}{\sigma^2}\right)\right]&=\frac{1}{\lambda_0}\int_0^{\infty}\mathrm{log}_2\left(\frac{px}{\sigma^2}\right)e^{-\frac{x}{\lambda_0}}dx=\mathrm{log}_2\left(\frac{p}{\sigma^2}\right)- E_0\mathrm{log}_2 e,\tag{A7} \label{expectation_first}
\end{align}
where $E_0$ is the Euler-Mascheroni constant. We then derive the distribution of $\Lambda$. The squared envelope $|h_k|^2$ and $|g_k|^2$ follows an exponential distribution and their PDF are given by $f_{|h_k|^2}(x)=\frac{1}{\lambda_h}e^{-\frac{x}{\lambda_h}}$ and $f_{|g_k|^2}(x)=\frac{1}{\lambda_g}e^{-\frac{x}{\lambda_g}}$. Thus, the PDF of $Z_k$ is
\begin{align}
    f_{Z_k}(z)&=\int_0^{\infty}\frac{1}{u}f_{|g_k|^2\cdot |h_k|^2}\left(\frac{z}{u},u\right)du=\int_0^{\infty}\frac{1}{u}f_{|g_k|^2}\left(\frac{z}{u}\right)f_{|h_k|^2}\left(u\right)du=\frac{1}{\lambda}\int_0^{\infty}\frac{1}{u}e^{-\frac{z}{\lambda_g u}-\frac{u}{\lambda_h}}du\nonumber\\
    &=\frac{2}{\lambda}K_0\left(2\sqrt{\frac{z}{\lambda}}\right),\label{f_z}\tag{A8}
\end{align}
where $K_0(\cdot)$ is the modified Bessel function of the second kind. The expectation and variance of $Z_k$ are given by
\begin{align}
   \mathbb{E}\left(Z_k\right)&=\int_0^{\infty}zf_{Z_k}(z)dz=\int_0^{\infty}z\frac{2}{\lambda}K_0\left(2\sqrt{\frac{z}{\lambda}}\right)dz \overset{(a)}{=} \frac{4}{\lambda} \int_0^{\infty} t^3K_0\left(\frac{2}{\sqrt{\lambda}}t\right)dt\overset{(b)}{=}\lambda,\tag{A9}\\
   \mathbb{V}\left(Z_k\right)&=\mathbb{E}\left(Z_k^2\right)-\mathbb{E}^2\left(Z_k\right)=\int_0^{\infty}z^2f_{Z_k}(z)dz-\lambda^2\overset{(c)}{=} \frac{4}{\lambda} \int_0^{\infty} t^5K_0\left(\frac{2}{\sqrt{\lambda}}t\right)dt-\lambda^2 \overset{(d)}{=}3\lambda^2,\tag{A10}
\end{align}
where $(a)$ and $(c)$ are obtained by denoting $t=\sqrt{z}$, and $(b)$ and $(d)$ follow from $\int_0^{\infty}x^{\mu}K_{\upsilon}\left(ax\right)dx=2^{\mu-1}a^{-\mu-1}\Gamma\left(\frac{1+\mu+\upsilon}{2}\right)\Gamma\left(\frac{1+\mu-\upsilon}{2}\right)$\cite{gradshteyn2014table}. Here, $\Gamma(x)$ is Gamma function and $\Gamma(x)=(x-1)!$ when $x$ is a positive integer. From the central limit theorem, we can obtain that for large $K$, $\Lambda$ approximately follows the Gaussian distribution ${\cal{N}}\left(K\lambda,3K\lambda^2\right)$. The PDF of $\Psi$ is thus given by
\begin{align}
    f_{\Psi}(\psi)=\frac{1}{\lambda_0\lambda\sqrt{6\pi K}}\int_0^{\infty}ue^{-\frac{\psi u}{\lambda_0}}e^{-\frac{\left(u-K\lambda\right)^2}{6K\lambda^2}}du. \tag{A11}
\end{align}
The second item in (\ref{eq_Eh}) is then calculated as
\begin{align}
    &\mathbb{E}_{\boldsymbol{h_0},\Lambda}\left[\mathrm{Ei}\left(-\frac{|h_0|^2}{\alpha^2 \Lambda}\right)\right]=\int_0^{\infty}\mathrm{Ei}\left(-\frac{\psi }{\alpha^2 }\right)f_{\Psi}\left(\psi\right)d\psi\nonumber\\
    &=\frac{1}{\lambda_0\lambda\sqrt{6\pi K}}\int_0^{\infty}\int_0^{\infty}\mathrm{Ei}\left(-\frac{\psi }{\alpha^2 }\right)ue^{-\frac{\psi u}{\lambda_0}}e^{-\frac{\left(u-K\lambda\right)^2}{6K\lambda^2}}du d\psi.\tag{A12}
\end{align}
Exchange the integration order and we obtain
\begin{align}
    &\mathbb{E}_{\boldsymbol{h_0},\Lambda}\left[\mathrm{Ei}\left(-\frac{|h_0|^2}{\alpha^2 \Lambda}\right)\right]=\frac{1}{\lambda_0\lambda\sqrt{6\pi K}}\int_0^{\infty}u\int_0^{\infty}\mathrm{Ei}\left(-\frac{\psi }{\alpha^2 }\right)e^{-\frac{\psi u}{\lambda_0}}d\psi e^{-\frac{\left(u-K\lambda\right)^2}{6K\lambda^2}}du. \tag{A13}\label{eqE1}
\end{align}
From \cite{gradshteyn2014table}, we have
\begin{align}
    \int_0^{\infty}\mathrm{Ei}\left(-\frac{\psi }{\alpha^2 }\right)e^{-\frac{\psi u}{\lambda_0}}d\psi=-\frac{\lambda_0}{u}\mathrm{ln}\left(1+\frac{\alpha^2 u}{\lambda_0}\right).\tag{A14}\label{lemm_Ei}
\end{align}
Substituting (\ref{lemm_Ei}) into (\ref{eqE1}), we have
\begin{align}
    \mathbb{E}_{\boldsymbol{h_0},\Lambda}\left[\mathrm{Ei}\left(-\frac{|h_0|^2}{\alpha^2 \Lambda}\right)\right]&=-\frac{1}{\lambda\sqrt{6\pi K}}\int_0^{\infty}\mathrm{ln}\left(1+\frac{\alpha^2 u}{\lambda_0}\right)e^{-\frac{\left(u-K\lambda\right)^2}{6K\lambda^2}}du\nonumber \\
    &=-\mathbb{E}_{\Lambda}\left[\mathrm{ln}\left(1+\frac{\alpha^2 \Lambda}{\lambda_0}\right)\right].\tag{A15} \label{expectation_second}
\end{align}
By using a second-order Taylor expansion with a Gaussian distribution input\cite{teh2006collapsed}, we have
\begin{align}
    \mathbb{E}\left[\mathrm{ln}\left(1+x\right)\right]\approx \mathrm{ln}\left(1+\mathbb{E}\left[x\right]\right)-\frac{\mathbb{V}[x]}{2\left(1+\mathbb{E}\left[x\right]\right)^2}.\tag{A16}\label{Taylor_expansion}
\end{align}
Thus, we obtain
\begin{align}
    &\mathbb{E}_{\Lambda}\left[\mathrm{ln}\left(1+\frac{\alpha^2 \Lambda}{\lambda_0}\right)\right]=\mathrm{ln}\left(1+\frac{K\alpha^2\lambda}{\lambda_0}\right)-\frac{3K\lambda^2\alpha^4}{2\left(\lambda_0+K\alpha^2\lambda\right)^2}.\tag{A17}\label{Gaussian_appro}
\end{align}
As a consequence, combining (\ref{expectation_first}), (\ref{expectation_second}) and (\ref{Gaussian_appro}), the ergodic rate of celullar transmission is given by
\begin{align}
    \mathbb{E}_{\boldsymbol{h}}\left[R_s^{SA}\right]=&\mathrm{log}_2\left(\frac{p}{\sigma^2}\right)+\bigg[\mathrm{ln}\left(1+\frac{K\alpha^2\lambda}{\lambda_0}\right)   -\frac{3K\lambda^2\alpha^4}{2\left(\lambda_0+K\alpha^2\lambda\right)^2}-E_0\bigg]\mathrm{log}_2 e.\tag{A18}
\end{align}

\section{Proof of Lemma 1}\label{proof of lemma1}
From the property of Bessel function, we have
\begin{align}
   \left(1-2\zeta\sqrt{ x}K_1\left(2\zeta\sqrt{ x}\right)\right)^{K-1}\rightarrow 1, \mathrm{as}\ x\rightarrow \infty,\tag{A19}\\
   K_0\left(2\zeta\sqrt{ x}\right) \rightarrow  \sqrt{\frac{\pi}{4\zeta\sqrt{x}}}e^{-2\zeta\sqrt{x}}, \mathrm{as} \ x\rightarrow \infty. \tag{A20}
    %\underset{x \rightarrow \infty}{\mathrm{lim}}\left(1-2\zeta\sqrt{ x}K_1\left(2\zeta\sqrt{ x}\right)\right)^{K-1}=1,\\
    %\underset{x \rightarrow \infty}{\mathrm{lim}}K_0\left(2\zeta\sqrt{ x}\right)=\sqrt{\frac{\pi}{4\zeta\sqrt{x}}}e^{-2\zeta\sqrt{x}}.
\end{align}
Thus, we can obtain
\begin{align}
&\underset{x \rightarrow \infty}{\mathrm{lim}}\mathrm{ln}\left(1+\epsilon x\right)\left(1-2\zeta\sqrt{ x}K_1\left(2\zeta\sqrt{ x}\right)\right)^{K-1}K_0\left(2\zeta\sqrt{ x}\right)\nonumber\\
&=\underset{x \rightarrow \infty}{\mathrm{lim}}\mathrm{ln}\left(\epsilon x\right)\sqrt{\frac{\pi}{4\zeta\sqrt{x}}}e^{-2\zeta\sqrt{x}}=\underset{\eta \rightarrow \infty}{\mathrm{lim}}\mathrm{ln}\left(\epsilon \eta^2\right)\sqrt{\frac{\pi}{4\zeta\eta}}e^{-2\zeta\eta}\nonumber\\
&=\underset{\eta \rightarrow \infty}{\mathrm{lim}}\frac{2\mathrm{ln}\left(\sqrt{\epsilon} \eta\right)}{e^{2\zeta\eta}}\sqrt{\frac{\pi}{4\zeta\eta}}.\tag{A21}
\end{align}
Since $\underset{\eta \rightarrow \infty}{\mathrm{lim}}\frac{2\mathrm{ln}\left(\sqrt{\epsilon} \eta\right)}{e^{2\zeta\eta}}=0$ and $\underset{\eta \rightarrow \infty}{\mathrm{lim}}\sqrt{\frac{\pi}{4\zeta\eta}}=0$, we can conclude that $\underset{x \rightarrow \infty}{\mathrm{lim}}f(x)=0$, which completes the proof.

\section{Proof of Theorem 2}\label{proof of rate s for SDA}
From (\ref{f_z}), the cumulative distribution function (CDF) of $Z_k$ is given by
\begin{align}
    F_{Z_k}(z)=1-2\sqrt{\frac{z}{\lambda}}K_1\left(2\sqrt{\frac{z}{\lambda}}\right).\label{F_z}\tag{A22}
\end{align}
We can obtain that the CDF and PDF of $Z$ are written as
\begin{align}
    F_{Z}(z)&=\left(F_{Z_k}(z)\right)^K=\left(1-2\sqrt{\frac{z}{\lambda}}K_1\left(2\sqrt{\frac{z}{\lambda}}\right)\right)^K,\tag{A23}\\
    f_{Z}(z)&=K\left(F_{Z_k}(z)\right)^{K-1}f_{Z_k}(z)=\frac{2K}{\lambda}\left(1-2\sqrt{\frac{z}{\lambda}}K_1\left(2\sqrt{\frac{z}{\lambda}}\right)\right)^{K-1} K_0\left(2\sqrt{\frac{z}{\lambda}}\right).\label{f_Z}\tag{A24}
\end{align}
Denote $\Theta=\frac{H}{Z}$. The PDF of $\Theta$ is given by
\begin{align}
    f_{\Theta}\left(\theta\right)&=\int_0^{\infty}uf_H(\theta u)f_Z(u)du\nonumber\\
    &=\frac{2K}{\lambda_0\lambda}\int_0^{\infty}ue^{-\frac{\theta u}{\lambda_0}}\left(1-2\sqrt{\frac{u}{\lambda}}K_1\left(2\sqrt{\frac{u}{\lambda}}\right)\right)^{K-1} K_0\left(2\sqrt{\frac{u}{\lambda}}\right)du.\tag{A25}
\end{align}
For the second item, the integration can be calculated as
\begin{align}
    &\mathbb{E}_{\boldsymbol{h_0},Z}\left[\mathrm{Ei}\left(-\frac{|h_0|^2}{\alpha^2 Z}\right)\right]=\int_0^{\infty}\mathrm{Ei}\left(-\frac{\theta }{\alpha^2 }\right)f_{\Theta}\left(\theta\right)d\theta\nonumber\\
    &=\frac{2K}{\lambda_0\lambda}\int_0^{\infty}\mathrm{Ei}\left(-\frac{\theta }{\alpha^2 }\right)\int_0^{\infty}ue^{-\frac{\theta u}{\lambda_0}} \left(1-2\sqrt{\frac{u}{\lambda}}K_1\left(2\sqrt{\frac{u}{\lambda}}\right)\right)^{K-1}K_0\left(2\sqrt{\frac{u}{\lambda}}\right)dud\theta.\tag{A26}
\end{align}
Change the integration order and we can obtain
\begin{align}
    &\mathbb{E}_{\boldsymbol{h_0},Z}\left[\mathrm{Ei}\left(-\frac{|h_0|^2}{\alpha^2 Z}\right)\right]\nonumber \\
    &=\frac{2K}{\lambda_0\lambda}\int_0^{\infty}\left[\int_0^{\infty}\mathrm{Ei}\left(-\frac{\theta }{\alpha^2 }\right)e^{-\frac{\theta u}{\lambda_0}}d\theta\right] u\left(1-2\sqrt{\frac{u}{\lambda}}K_1\left(2\sqrt{\frac{u}{\lambda}}\right)\right)^{K-1}K_0\left(2\sqrt{\frac{u}{\lambda}}\right)du.\tag{A27}\label{E_diversity}
\end{align}
Substituting (\ref{lemm_Ei}) into (\ref{E_diversity}), we can obtain
\begin{align}
    &\mathbb{E}_{\boldsymbol{h_0},Z}\left[\mathrm{Ei}\left(-\frac{|h_0|^2}{\alpha^2 Z}\right)\right]\nonumber \\
    &=-\frac{2K}{\lambda}\int_0^{\infty}\mathrm{ln}\left(1+\frac{\alpha^2 u}{\lambda_0}\right)\left(1-2\sqrt{\frac{u}{\lambda}}K_1\left(2\sqrt{\frac{u}{\lambda}}\right)\right)^{K-1}K_0\left(2\sqrt{\frac{u}{\lambda}}\right)du.\tag{A28}\label{Second_item}
\end{align}
%From (\ref{Second_item}), we find it difficult to obtain an accurate closed-form expression and thus we provide the following proposition for numerical analysis.
According to Lemma~\ref{pro3}, we can transform the infinite integral in (\ref{Second_item}) into the following finite integral form:
\begin{align}
    &\mathbb{E}_{\boldsymbol{h_0},Z}\left[\mathrm{Ei}\left(-\frac{|h_0|^2}{\alpha^2 Z}\right)\right]\nonumber \\
    &\approx -\frac{2K}{\lambda}\int_0^{M_1}\mathrm{ln}\left(1+\frac{\alpha^2 u}{\lambda_0}\right) \left(1-2\sqrt{\frac{u}{\lambda}}K_1\left(2\sqrt{\frac{u}{\lambda}}\right)\right)^{K-1}K_0\left(2\sqrt{\frac{u}{\lambda}}\right)du,\tag{A29}
\end{align}
where $M_1$ is a large number. By applying Gaussian-Chebyshev quadrature\cite{hildebrand1987introduction}, we can obtain
\begin{align}
    \mathbb{E}_{\boldsymbol{h_0},Z}\left[\mathrm{Ei}\left(-\frac{|h_0|^2}{\alpha^2 Z}\right)\right]\approx -\frac{KM_1\pi}{\lambda n}\sum_{i=1}^{n}l\left(\phi_i\right),\tag{A30}
\end{align}
where $\phi_i=\mathrm{cos}\left(\frac{2i-1}{2n}\pi\right)$, $n$ is a complexity-accuracy tradeoff parameter and
\begin{align}
l(x) &=\mathrm{ln}\left(1+\frac{\alpha^2 \left(M_1x+M_1\right)}{2\lambda_0}\right)\Bigg(1-2\sqrt{\frac{M_1x+M_1}{2\lambda}} K_1\left(2\sqrt{\frac{M_1x+M_1}{2\lambda}}\right)\Bigg)^{K-1}\nonumber \\
&\times K_0\left(2\sqrt{\frac{M_1x+M_1}{2\lambda}}\right) \times \sqrt{1-x^2}.\tag{A31}
\end{align}
The ergodic rate for decoding $s(n)$ is thus given by
\begin{align}
    \mathbb{E}_{\boldsymbol{h}}\left[R_s^{SDA}\right] \approx \mathrm{log}_2\left(\frac{p}{\sigma^2}\right)- E_0\mathrm{log}_2 e+\frac{KM_1\pi\mathrm{log}_2 e}{\lambda n}\sum_{i=1}^{n}l\left(\phi_i\right).\tag{A32}
\end{align}

\section{Proof of Theorem 3}\label{proof of outage}
The outage probability is transformed to
\begin{align}
P_{out,r}^{SA}&=P\left\{\frac{H}{\alpha^2\Lambda}\mathrm{log}_2 e+\mathrm{log}_2\left( \frac{p\alpha^2\Lambda}{\sigma^2}\right)-E_0\mathrm{log}_2 e \leq \tilde{R}_s\right\}.\tag{A33}\label{out_trans}
\end{align}
Denoting $a_1=\frac{\mathrm{log}_2 e}{\alpha^2}$ and $a_2=\tilde{R}_s-\mathrm{log}_2\left( \frac{p\alpha^2}{\sigma^2}\right)+E_0\mathrm{log}_2 e$, we further obtain
\begin{align}
    P_{out,r}^{SA}&=P\left\{H\leq \frac{\Lambda\left(a_2-\mathrm{log}_2 \Lambda\right)}{a_1}\right\}=\int_{0}^{2^{a_2}}\int_{0}^{\frac{y\left(a_2-\mathrm{log}_2 y\right)}{a_1}}f_{H}(x) f_{\Lambda}(y)dx dy.\tag{A34}\label{P_SA_integration}
\end{align}
Since the Gaussian approximation is only accurate for large values of $K$ and the outage analysis requires a more accurate approximation, we use an RV with a generalized-K distribution to approximate $\Lambda$\cite{chatzidiamantis2009distribution,al2010approximation}. Note that certain research, such as~\cite{zhao2020backscatter}, have considered a Gamma distribution $Gamma\left(\beta_1,\beta_2\right)$ for approximation when $K$ is not large. The expectation and variance of $\Lambda$ are used to calculate the parameters $\beta_1$ and $\beta_2$. Nonetheless, the simulations in Section~\ref{sec-simulation} imply that, when compared to the Gaussian and Gamma distributions, the generalized-K distribution is the most accurate approximation for $\Lambda$.

Denote $\hat{\Lambda}$ as the approximate RV with a three-parameter generalized-K distribution for $\Lambda$. The PDF of $\hat{\Lambda}$ is given by
\begin{align}
    f_{\hat{\Lambda}}(x;m_{1,\Lambda},m_{2,\Lambda},\Omega_{\Lambda})&=\frac{2\left(m_{1,\Lambda}m_{2,\Lambda}\right)^{\frac{m_{1,\Lambda}+m_{2,\Lambda}}{2}}x^{\frac{m_{1,\Lambda}+m_{2,\Lambda}}{2}-1}}{\Gamma(m_{1,\Lambda})\Gamma(m_{2,\Lambda})\Omega_{\Lambda}^{\frac{m_{1,\Lambda}+m_{2,\Lambda}}{2}}} K_{m_{1,\Lambda}-m_{2,\Lambda}}\left[2\left(\frac{m_{1,\Lambda}m_{2,\Lambda}}{\Omega_{\Lambda}}x\right)^{\frac{1}{2}}\right],\tag{A35}\label{PDF_KG}
\end{align}
where $m_{1,\Lambda} \geq 0$ and $m_{2,\Lambda} \geq 0$ are the distribution shaping parameters, and $\Omega_{\Lambda}=\mathbb{E} (\hat{\Lambda})$. Its CDF is given by
\begin{align}
    F_{\hat{\Lambda}}(x)=\frac{1}{\Gamma(m_{1,\Lambda})\Gamma(m_{2,\Lambda})}
\MeijerG*{2}{1}{1}{3}{1}{m_{1,\Lambda},m_{2,\Lambda},0}{\frac{m_{1,\Lambda}m_{2,\Lambda}}{\Omega_{\Lambda}}x},\tag{A36}\label{KG_CDF}
\end{align}
where $G\left[\cdot \right]$ is Meijer's G function.
Alternatively, the parameter $m_{1,\Lambda}$ can take the values of $K$, $\frac{K+\sqrt{K^2+3K}}{3}$, $K+\epsilon_{\Lambda}$ ($\epsilon_{\Lambda}$ is a adjustment parameter), depending on the approximation accuracy \cite{al2010approximation,chatzidiamantis2011distribution}. In this paper, we choose $m_{1,\Lambda}=\frac{K+\sqrt{K^2+3K}}{3}$ since it achieves the best approximation for outage analysis. Specifically, the parameters for $\Lambda$ are given by
\begin{align}
    m_{1,\Lambda}=m_{2,\Lambda}=\frac{K+\sqrt{K^2+3K}}{3},\Omega_{\Lambda}=K\lambda.\tag{A37}
\end{align}
By substituting $f_H(x)=\frac{1}{\lambda_0}e^{-\frac{1}{\lambda_0}x}$ into (\ref{P_SA_integration}), we have
\begin{align}
    P_{out,r}^{SA}&=\int_{0}^{2^{a_2}}\left[1-e^{-\frac{y\left(a_2-\mathrm{log}_2 y\right)}{a_1\lambda_0}}\right]f_{\hat{\Lambda}}(y)dy=F_{\hat{\Lambda}}(2^{a_2})-B_1.\tag{A38}
\end{align}
Here, applying Gaussian-Chebyshev quadrature, $B_1$ is derived as
\begin{align}
    B_1=\int_{0}^{2^{a_2}}e^{-\frac{y\left(a_2-\mathrm{log}_2 y\right)}{a_1\lambda_0}}f_{\hat{\Lambda}}(y)dy=\frac{ \pi a_32^{a_2}}{n}\sum_{i=1}^{n}o\left(\phi_i\right),\tag{A39}
\end{align}
where $a_3=\frac{m_{1,\Lambda}^{2m_{1,\Lambda}}}{\Gamma^2(m_{1,\Lambda}){(K\lambda)}^{m_{1,\Lambda}}}$ and $o(x)=e^{-\frac{x\left(a_2-\mathrm{log}_2 x\right)}{a_1\lambda_0}}x^{m_{1,\Lambda}-1}K_0\left(2\sqrt{\frac{m_{1,\Lambda}^2x}{K\lambda}}\right)\sqrt{1-x^2}$.

\bibliography{reference}

\begin{thebibliography}{10}
\providecommand{\url}[1]{#1}
\csname url@samestyle\endcsname
\providecommand{\newblock}{\relax}
\providecommand{\bibinfo}[2]{#2}
\providecommand{\BIBentrySTDinterwordspacing}{\spaceskip=0pt\relax}
\providecommand{\BIBentryALTinterwordstretchfactor}{4}
\providecommand{\BIBentryALTinterwordspacing}{\spaceskip=\fontdimen2\font plus
\BIBentryALTinterwordstretchfactor\fontdimen3\font minus
  \fontdimen4\font\relax}
\providecommand{\BIBforeignlanguage}[2]{{%
\expandafter\ifx\csname l@#1\endcsname\relax
\typeout{** WARNING: IEEEtran.bst: No hyphenation pattern has been}%
\typeout{** loaded for the language `#1'. Using the pattern for}%
\typeout{** the default language instead.}%
\else
\language=\csname l@#1\endcsname
\fi
#2}}
\providecommand{\BIBdecl}{\relax}
\BIBdecl

\bibitem{wang2022multiple}
J.~Wang, X.~Ding, Q.~Zhang, and Y.-C. Liang, ``Multiple access for symbiotic
  radios: Facilitating massive {IoT} connections with cellular networks,'' in
  \emph{Proc. IEEE Global Commun. Conf. (GLOBECOM)}.\hskip 1em plus 0.5em minus
  0.4em\relax IEEE, Dec. 2022, pp. 1--6.

\bibitem{matti2019key}
M.~Latva-aho and K.~Leppanen, ``Key drivers and research challenges for 6{G}
  ubiquitous wireless intelligence,'' \emph{University of Oulu, White Paper},
  2019. Available: http://urn.fi/urn:isbn:9789526223544.

\bibitem{xiaohutowards}
X.~You \emph{et~al.}, ``Towards {6G} wireless communication networks: Vision,
  enabling technologies, and new paradigm shifts,'' \emph{Sci. China Inf.
  Sci.}, vol.~64, no.~1, pp. 1--74, 2021.

\bibitem{guo2021enabling}
F.~Guo, F.~R. Yu, H.~Zhang, X.~Li, H.~Ji, and V.~C. Leung, ``Enabling massive
  {IoT} toward {6G}: A comprehensive survey,'' \emph{IEEE Internet Things J.},
  2021.

\bibitem{nguyen20216g}
D.~C. Nguyen, M.~Ding, P.~N. Pathirana, A.~Seneviratne, J.~Li, D.~Niyato,
  O.~Dobre, and H.~V. Poor, ``{6G} internet of things: A comprehensive
  survey,'' \emph{IEEE Internet Things J.}, 2021.

\bibitem{Ericsson2021}
\BIBentryALTinterwordspacing
{Ericsson Public Information}, ``Ericsson mobility report {Q4} 2021 update,''
  \emph{Technique Report}, 2021. [Online]. Available:
  \url{https://www.ericsson.com/4ad7e9/assets/local/reports-papers/mobility-report/documents/2021/ericsson-mobility-report-november-2021.pdf}
\BIBentrySTDinterwordspacing

\bibitem{long2019symbiotic}
R.~Long, Y.-C. Liang, H.~Guo, G.~Yang, and R.~Zhang, ``Symbiotic radio: A new
  communication paradigm for passive {Internet of Things},'' \emph{IEEE
  Internet Things J.}, vol.~7, no.~2, pp. 1350--1363, 2019.

\bibitem{liang2020symbiotic}
Y.-C. Liang, Q.~Zhang, E.~G. Larsson, and G.~Y. Li, ``Symbiotic radio:
  Cognitive backscattering communications for future wireless networks,''
  \emph{IEEE Trans. Cogn. Commun. Netw.}, vol.~6, no.~4, pp. 1242--1255, 2020.

\bibitem{liang2022symbiotic}
Y.-C. Liang, R.~Long, Q.~Zhang, and D.~Niyato, ``Symbiotic communications:
  Where marconi meets darwin,'' \emph{IEEE Wireless Commun.}, vol.~29, no.~1,
  pp. 144--150, 2022.

\bibitem{imoize20216g}
A.~L. Imoize, O.~Adedeji, N.~Tandiya, and S.~Shetty, ``{6G} enabled smart
  infrastructure for sustainable society: Opportunities, challenges, and
  research roadmap,'' \emph{Sensors}, vol.~21, no.~5, p. 1709, 2021.

\bibitem{yazar20206g}
A.~Yazar, S.~Dogan-Tusha, and H.~Arslan, ``{6G} vision: An ultra-flexible
  perspective,'' \emph{ITU J. Future Evolving Technol.}, vol.~1, no.~1, pp.
  121--140, 2020.

\bibitem{chen2020vision}
S.~Chen, Y.-C. Liang, S.~Sun, S.~Kang, W.~Cheng, and M.~Peng, ``Vision,
  requirements, and technology trend of {6G}: How to tackle the challenges of
  system coverage, capacity, user data-rate and movement speed,'' \emph{IEEE
  Wireless Commun.}, vol.~27, no.~2, pp. 218--228, 2020.

\bibitem{liu2018backscatter}
W.~Liu, Y.-C. Liang, Y.~Li, and B.~Vucetic, ``Backscatter multiplicative
  multiple-access systems: Fundamental limits and practical design,''
  \emph{IEEE Trans. Wireless Commun.}, vol.~17, no.~9, pp. 5713--5728, 2018.

\bibitem{guo2021capacity}
Y.~Guo, G.~Wang, R.~Xu, R.~He, X.~Wei, and C.~Tellambura, ``Capacity analysis
  for wireless symbiotic communication systems with {BPSK} tags under
  sensitivity constraint,'' \emph{IEEE Commun. Lett.}, vol.~26, no.~1, pp.
  44--48, 2021.

\bibitem{zhou2019ergodic}
S.~Zhou, W.~Xu, K.~Wang, C.~Pan, M.-S. Alouini, and A.~Nallanathan, ``Ergodic
  rate analysis of cooperative ambient backscatter communication,'' \emph{IEEE
  Wireless Commun. Lett.}, vol.~8, no.~6, pp. 1679--1682, 2019.

\bibitem{darsena2017modeling}
D.~Darsena, G.~Gelli, and F.~Verde, ``Modeling and performance analysis of
  wireless networks with ambient backscatter devices,'' \emph{IEEE Trans.
  Commun.}, vol.~65, no.~4, pp. 1797--1814, 2017.

\bibitem{zhao2018ambient}
W.~Zhao, G.~Wang, R.~Fan, L.-S. Fan, and S.~Atapattu, ``Ambient backscatter
  communication systems: Capacity and outage performance analysis,'' \emph{IEEE
  Access}, vol.~6, pp. 22\,695--22\,704, 2018.

\bibitem{zhang2019backscatter}
Q.~Zhang, L.~Zhang, Y.-C. Liang, and P.-Y. Kam, ``Backscatter-{NOMA}: A
  symbiotic system of cellular and {Internet-of-Things} networks,'' \emph{IEEE
  Access}, vol.~7, pp. 20\,000--20\,013, 2019.

\bibitem{yang2021energy}
H.~Yang, Y.~Ye, K.~Liang, and X.~Chu, ``Energy efficiency maximization for
  symbiotic radio networks with multiple backscatter devices,'' \emph{IEEE Open
  J. Commun. Soc.}, vol.~2, pp. 1431--1444, 2021.

\bibitem{han2021design}
S.~Han, Y.-C. Liang, and G.~Sun, ``The design and optimization of random code
  assisted {multi-BD} symbiotic radio system,'' \emph{IEEE Trans. Wireless
  Commun.}, vol.~20, no.~8, pp. 5159--5170, 2021.

\bibitem{liu2020symbiotic}
Y.~Liu, P.~Ren, and Q.~Du, ``Symbiotic communication: Concurrent transmission
  for multi-users based on backscatter communication,'' in \emph{Proc.
  International Conference on Wireless Communications and Signal Processing
  (WCSP)}.\hskip 1em plus 0.5em minus 0.4em\relax IEEE, 2020, pp. 835--839.

\bibitem{chen2020stochastic}
X.~Chen, H.~V. Cheng, K.~Shen, A.~Liu, and M.-J. Zhao, ``Stochastic transceiver
  optimization in multi-tags symbiotic radio systems,'' \emph{IEEE Internet
  Things J.}, vol.~7, no.~9, pp. 9144--9157, 2020.

\bibitem{ding2021advantages}
Z.~Ding and H.~V. Poor, ``Advantages of {NOMA} for multi-user backcom
  networks,'' \emph{IEEE Commun. Lett.}, vol.~25, no.~10, pp. 3408--3412, 2021.

\bibitem{zhang2021minimum}
R.~Zhang, X.~Kang, and Y.-C. Liang, ``Minimum throughput maximization for
  peer-assisted {NOMA}-plus-{TDMA} symbiotic radio networks,'' \emph{IEEE
  Wireless Commun. Lett.}, vol.~10, no.~9, pp. 1847--1851, 2021.

\bibitem{mishra2019sum}
D.~Mishra and E.~G. Larsson, ``Sum throughput maximization in multi-tag
  backscattering to multiantenna reader,'' \emph{IEEE Trans. Commun.}, vol.~67,
  no.~8, pp. 5689--5705, 2019.

\bibitem{chen2015interference}
X.~Chen and C.~Yuen, ``On interference alignment with imperfect {CSI}:
  Characterizations of outage probability, ergodic rate and {SER},'' \emph{IEEE
  Trans. Veh. Technol.}, vol.~65, no.~1, pp. 47--58, 2015.

\bibitem{shin2003capacity}
H.~Shin and J.~H. Lee, ``Capacity of multiple-antenna fading channels: Spatial
  fading correlation, double scattering, and keyhole,'' \emph{IEEE Trans. Inf.
  Theory}, vol.~49, no.~10, pp. 2636--2647, 2003.

\bibitem{jeon2011bounds}
H.~Jeon, N.~Kim, J.~Choi, H.~Lee, and J.~Ha, ``Bounds on secrecy capacity over
  correlated ergodic fading channels at high {SNR},'' \emph{IEEE Trans. Inf.
  Theory}, vol.~57, no.~4, pp. 1975--1983, 2011.

\bibitem{viswanath2002opportunistic}
P.~Viswanath, D.~N.~C. Tse, and R.~Laroia, ``Opportunistic beamforming using
  dumb antennas,'' \emph{IEEE Trans. Inf. Theory}, vol.~48, no.~6, pp.
  1277--1294, 2002.

\bibitem{al2020performance}
Y.~H. Al-Badarneh, M.-S. Alouini, and C.~N. Georghiades, ``Performance analysis
  of monostatic multi-tag backscatter systems with general order tag
  selection,'' \emph{IEEE Wireless Commun. Lett.}, vol.~9, no.~8, pp.
  1201--1205, 2020.

\bibitem{david2004order}
H.~A. David and H.~N. Nagaraja, \emph{Order statistics}.\hskip 1em plus 0.5em
  minus 0.4em\relax John Wiley \& Sons, 2004.

\bibitem{jin2014ergodic}
S.~Jin, X.~Liang, K.-K. Wong, X.~Gao, and Q.~Zhu, ``Ergodic rate analysis for
  multipair massive {MIMO} two-way relay networks,'' \emph{IEEE Trans. Wireless
  Commun.}, vol.~14, no.~3, pp. 1480--1491, 2014.

\bibitem{song2006asymptotic}
G.~Song and Y.~Li, ``Asymptotic throughput analysis for channel-aware
  scheduling,'' \emph{IEEE Trans. Commun.}, vol.~54, no.~10, pp. 1827--1834,
  2006.

\bibitem{li2019capacity}
D.~Li, W.~Peng, and F.~Hu, ``Capacity of backscatter communication systems with
  tag selection,'' \emph{IEEE Trans. Veh. Technol.}, vol.~68, no.~10, pp.
  10\,311--10\,314, 2019.

\bibitem{yang2020outage}
L.~Yang, Y.~Yang, D.~B. da~Costa, and I.~Trigui, ``Outage probability and
  capacity scaling law of multiple {RIS}-aided networks,'' \emph{IEEE Wireless
  Commun. Lett.}, vol.~10, no.~2, pp. 256--260, 2020.

\bibitem{wang2003simple}
Z.~Wang and G.~B. Giannakis, ``A simple and general parameterization
  quantifying performance in fading channels,'' \emph{IEEE Trans. Commun.},
  vol.~51, no.~8, pp. 1389--1398, 2003.

\bibitem{rappaport2010wireless}
T.~S. Rappaport, \emph{Wireless communications: Principles and practice, 2nd
  ed}.\hskip 1em plus 0.5em minus 0.4em\relax Upper Saddle River, NJ, USA:
  Prentice-Hall, 2001.

\bibitem{griffin2009complete}
J.~D. Griffin and G.~D. Durgin, ``Complete link budgets for backscatter-radio
  and {RFID} systems,'' \emph{IEEE Antennas Propag. Mag.}, vol.~51, no.~2, pp.
  11--25, 2009.

\bibitem{tse2005fundamentals}
D.~Tse and P.~Viswanath, \emph{Fundamentals of wireless communication}.\hskip
  1em plus 0.5em minus 0.4em\relax Cambridge university press, 2005.

\bibitem{gradshteyn2014table}
I.~S. Gradshteyn and I.~M. Ryzhik, \emph{Table of integrals, series, and
  products}.\hskip 1em plus 0.5em minus 0.4em\relax Academic press, 2014.

\bibitem{teh2006collapsed}
Y.~Teh, D.~Newman, and M.~Welling, ``A collapsed variational {Bayesian}
  inference algorithm for latent {Dirichlet} allocation,'' \emph{Adv. Neural
  Inf. Process. Syst.}, vol.~19, 2006.

\bibitem{hildebrand1987introduction}
F.~B. Hildebrand, \emph{Introduction to numerical analysis}.\hskip 1em plus
  0.5em minus 0.4em\relax Courier Corporation, 1987.

\bibitem{chatzidiamantis2009distribution}
N.~D. Chatzidiamantis, G.~K. Karagiannidis, and D.~S. Michalopoulos, ``On the
  distribution of the sum of gamma-gamma variates and application in {MIMO}
  optical wireless systems,'' in \emph{Proc. IEEE Global Commun. Conf.
  (GLOBECOM)}.\hskip 1em plus 0.5em minus 0.4em\relax IEEE, 2009, pp. 1--6.

\bibitem{al2010approximation}
S.~Al-Ahmadi and H.~Yanikomeroglu, ``On the approximation of the {PDF} of the
  sum of independent {generalized-K} {RVs} by another {generalized-K} {PDF}
  with applications to distributed antenna systems,'' in \emph{Proc. IEEE
  Wireless Commun. Netw. Conf. (WCNC)}.\hskip 1em plus 0.5em minus 0.4em\relax
  IEEE, 2010, pp. 1--6.

\bibitem{zhao2020backscatter}
W.~Zhao, G.~Wang, S.~Atapattu, T.~A. Tsiftsis, and C.~Tellambura, ``Is
  backscatter link stronger than direct link in reconfigurable intelligent
  surface-assisted system?'' \emph{IEEE Commun. Lett.}, vol.~24, no.~6, pp.
  1342--1346, 2020.

\bibitem{chatzidiamantis2011distribution}
N.~D. Chatzidiamantis and G.~K. Karagiannidis, ``On the distribution of the sum
  of gamma-gamma variates and applications in {RF} and optical wireless
  communications,'' \emph{IEEE Trans. Commun.}, vol.~59, no.~5, pp. 1298--1308,
  2011.

\end{thebibliography}
\bibliographystyle{IEEEtran}
\end{document}